\documentclass[12pt,preprint]{aastex}
\usepackage{natbib}
\usepackage[onecolumn,numberedappendix]{emulateapj5}
\usepackage{epsfig}
\usepackage{rotating}
\tightenlines
\slugcomment{Accepted to the Astrophysical Journal}

\newcommand \lsim{\mathrel{\rlap{\lower4pt\hbox{\hskip1pt$\sim$}}
    \raise1pt\hbox{$<$}}}
\newcommand \gsim{\mathrel{\rlap{\lower4pt\hbox{\hskip1pt$\sim$}}
    \raise1pt\hbox{$>$}}}

\newcommand{\beq}{\begin{equation}}
\newcommand{\eeq}{\end{equation}}
\newcommand{\beqa}{\begin{eqnarray}}
\newcommand{\eeqa}{\end{eqnarray}}

\newlength{\figwidth}
\begin{document}

\title{A Test of Star Formation Laws in Disk Galaxies. II.\\
Dependence on dynamical properties}


\author{Chutipong Suwannajak$^1$, Jonathan C. Tan$^{1,2}$, Adam K. Leroy$^3$}
\affil{$^1$Department of Astronomy, University of Florida, Gainesville, Florida 32611, USA}
\affil{$^2$Department of Physics, University of Florida, Gainesville, Florida 32611, USA}
\affil{$^3$National Radio Astronomy Observatory, 520 Edgemont Road, Charlottesville, Virginia 22903, USA}

\begin{abstract}
We use the observed radial profiles of the mass surface densities of
total, $\Sigma_g$, and molecular, $\Sigma_{\rm{H2}}$, gas, rotation
velocity and star formation rate (SFR) surface density,
$\Sigma_{\rm{sfr}}$, of the molecular-rich
($\Sigma_{\rm{H2}}\ge\Sigma_{\rm{HI}}/2$) regions of 16 nearby disk
galaxies to test several star formation laws: a ``Kennicutt-Schmidt''
law, $\Sigma_{\rm{sfr}}=A_g\Sigma_{g,2}^{1.5}$; a ``Constant
Molecular'' law, $\Sigma_{\rm sfr}=A_{\rm H2}\Sigma_{\rm{H2,2}}$; the
turbulence-regulated laws of Krumholz \& McKee (KM05) and Krumholz,
McKee \& Tumlinson (KMT09), a ``Gas-$\Omega$'' law,
$\Sigma_{\rm{sfr}}=B_\Omega\Sigma_g\Omega$; and a shear-driven ``GMC
Collision'' law,
$\Sigma_{\rm{sfr}}=B_{\rm{CC}}\Sigma_g\Omega(1-0.7\beta)$, where
$\beta\equiv\:d\:{\rm{ln}}\:v_{\rm{circ}}/d\:{\rm{ln}}\:r$.  If
allowed one free normalization parameter for each galaxy, these laws
predict the SFR with rms errors of factors of 1.4 to 1.8.  If a single
normalization parameter is used by each law for the entire galaxy
sample, then rms errors range from factors of 1.5 to 2.1.  Although
the Constant Molecular law gives the smallest rms errors, the
improvement over the KMT, Kennicutt-Schmidt and GMC Collision laws is
not especially significant, particularly given the different
observational inputs that the laws utilize and the scope of included
physics, which ranges from empirical relations to detailed treatment
of interstellar medium processes.  We next search for systematic
variation of star formation law parameters with local and global
galactic dynamical properties of disk shear rate (related to $\beta$),
rotation speed and presence of a bar.  We demonstrate with high
significance that higher shear rates enhance star formation efficiency
per local orbital time. Such a trend is expected if GMC collisions
play an important role in star formation, while an opposite trend
would be expected if development of disk gravitational instabilities
is the controlling physics.
\end{abstract}


\section{Introduction}\label{S:intro}

Understanding the rate at which stars form from a given galactic gas
inventory is a basic input for models of galaxy evolution. Global and
kiloparsec-scale correlations between star formation activity, gas
content and galactic dynamical properties have been observed (e.g.,
Kennicutt \& Evans 2012). However, most star formation is known to
occur in highly clustered $\sim 1-10$~pc-scale regions within giant
molecular clouds (GMCs) and the physical processes linking these large
and small scales, i.e., the ``micro-physics'' of galactic star
formation laws, remain uncertain.

Tan (2010, hereafter Paper I), analyzed data from Leroy et al. (2008)
for the molecular dominated regions of 12 nearby disk galaxies.  The
predictions of six star formation laws, described below, were tested
against the observed radial profiles in the galaxies. 

In this paper, after summarizing the star formation (SF) laws to be
considered (\S\ref{S:laws}) and adopting similar methods as Paper I
(\S\ref{S:method}), we have extended this work by: (1) utilizing a modestly
expanded sample of 16 galaxies, which are now explicity selected to be
relatively large disk galaxies with mean circular velocity $\geq
100\:{\rm km\:s^{-1}}$ (11 galaxies overlap with the sample of Paper I); (2) also considering ``molecular rich''
regions where $\Sigma_{\rm HI}/2<\Sigma_{\rm H2}<\Sigma_{\rm HI}$, in
addition to the ``molecular dominated'' regions (the results of
relative comparison of the different SF laws in these regions are
presented in \S\ref{S:lawtest}); (3) searching for correlations of SF
law parameters with galactic dynamical properties
(\S\ref{S:properties}), i.e., galactic disk shear (rotation curve
gradient), rotation speed, and presence of a bar.
We conclude in \S\ref{S:conclusion}.


\section{Overview of Star Formation ``Laws'' to be Tested}\label{S:laws}

Here we overview the various star formation ``laws'' that we will test
in this paper. These vary in their nature from being simple empirical
relations to being the predictions of more detailed models of physical
processes in galactic interstellar media. There are also varying
ranges of physical conditions over which these laws are expected to be
valid. We note also that the measurement of star formation rates and
gas masses, which are key ingredients in these laws suffer from
significant systematic, potentially correlated, uncertainties (see,
e.g., discussions in Leroy et al. 2008, 2013; Sandstrom et al. 2013).

Considering global disk-averages, Kennicutt (1998) presented an
empirical relation, hereafter the Kennicutt-Schmidt law, between the
disk plane surface density of star formation rate (SFR), $\Sigma_{\rm
  sfr}$, and the total gas mass surface density:
\begin{equation}
\label{sfr1}
\Sigma_{\rm sfr} = A_{g} \Sigma_{g,2}^{\alpha_{g}},
\end{equation}
where $A_g= 0.158 \pm 0.044 \:M_\odot\:{\rm yr^{-1}\:kpc^{-2}}$,
$\Sigma_{g,2} = \Sigma_g / 100 M_\odot {\rm pc^{-2}}$, and $\alpha_{g}
= 1.4\pm0.15$. The dynamic range of this relation covers from the
molecular-rich regions of normal galaxies to the molecular-dominated
regions in galactic centers and in starburst galaxies. 
Theoretical and numerical models that relate the SFR to the growth
rate of large scale gravitational instabilities in a disk predict
$\alpha_{g}\simeq 1.5$ (e.g. Larson 1988; Elmegreen 1994, 2002; Wang
\& Silk 1994; Li, Mac Low, \& Klessen 2006), as long as the gas scale
height does not vary much from galaxy to galaxy or, for a local form
of the relation, within the galaxy.

Alternatively, based on a study of 12 nearby disk galaxies resolved at
$\sim$ 1 kpc resolution, Leroy et al. (2008) (see also Bigiel et
al. 2008) concluded that
\begin{equation}
\label{sfrH2}
\Sigma_{\rm sfr} = A_{\rm H2} \Sigma_{\rm H2,2},
\end{equation}
where $A_{\rm H2}=(5.25 \pm 2.5) \times 10^{-2}\:M_\odot {\rm
  yr^{-1}\:kpc^{-2}}$ and $\Sigma_{\rm H2,2} = \Sigma_{\rm H2} / 100
M_\odot {\rm pc^{-2}}$. The values of $\Sigma_{\rm H2}$ covered a
range from $\sim 4 - 100 M_\odot\:{\rm pc}^{-2}$ and were estimated
assuming a constant ``X'' conversion factor of CO line emission to $\rm H_2$
column density.
Leroy et al. (2013) have shown that, from a similar study of 30 nearby disk
galaxies, $A_{\rm H2}= 4.5 \times
10^{-2}\:M_\odot\: {\rm yr^{-1}\:kpc^{-2}}$ or $\tau_{\rm dep}^{\rm
  H2} = \Sigma_{\rm H2}/ \Sigma_{\rm sfr} = A_{\rm H2}^{-1} = 2.2\: \rm Gyr$ with
$\approx$ 0.3 dex scatter.  This star formation relation will be
referred to as the Constant Molecular law.  

The turbulence-regulated star formation model of Krumholz $\&$ McKee
(2005) (hereafter the KM05 law) predicts galactic
star formation rates by assuming GMCs are virialized and that their
surfaces are in pressure equilibrium with the large scale interstellar
medium (ISM) pressure of a Toomre (1964) $Q\simeq 1.5$ disk. They
predict
\begin{equation}
\label{sfrKM}
\Sigma_{\rm sfr} = A_{\rm KM} f_{\rm GMC} \phi_{\bar{P},6}^{0.34} Q_{1.5}^{-1.32} \Omega_0^{1.32} \Sigma_{g,2}^{0.68},
\end{equation}
where $A_{\rm KM}=9.5 M_\odot \: {\rm yr^{-1}\:kpc^{-2}}$, $f_{\rm
  GMC}$ is the mass fraction of gas in GMCs, $\phi_{\bar{P},6}$ is the
ratio of the mean pressure in a GMC to the surface pressure here
normalized to a fiducial value of 6 but estimated to vary as
$\phi_{\bar{P}}=10-8f_{\rm GMC}$, $Q_{1.5}=Q/1.5$, and $\Omega_0$ is
$\Omega$, the orbital angular frequency, in units of $\rm
Myr^{-1}$. We assume that $f_{\rm GMC}= \Sigma_{\rm H2}/\Sigma_g$
based on resolved studies of GMC populations and molecular gas content
in the Milky Way and nearby galaxies (Solomon et al. 1987; Blitz et
al. 2007).

Krumholz, McKee \& Tumlinson (2009) (hereafter the
KMT09 law) presented a two component turbulence-regulated star formation law
\begin{eqnarray}
\label{sfrKMT}
\Sigma_{\rm sfr}  =  A_{\rm KMT} f_{\rm GMC} \Sigma_{g,2} \times \left\{ 
\begin{array}{lc}
\left(\Sigma_g/85 M_\odot {\rm pc^{-2}}\right)^{-0.33}, & \Sigma_g< 85\:M_\odot {\rm pc^{-2}}\\
\left(\Sigma_g/85 M_\odot {\rm pc^{-2}}\right)^{0.33}, & \Sigma_g> 85\:M_\odot {\rm pc^{-2}}
\end{array}
\right\}
\end{eqnarray}
where $A_{\rm KMT}=3.85 \times 10^{-2}\:M_\odot {\rm
  yr^{-1}\:kpc^{-2}}$. GMCs are assumed to be in pressure equilibrium
with the ISM only in the high $\Sigma_g$ regime. In the low regime,
GMCs are assumed to have constant internal pressures set by \ion{H}{2}
region feedback (Matzner 2002). Krumholz et al. (2012) presented a
test of this turbulence regulated star formation law against observations of
star formation from scales of individual GMCs to entire galaxies.

K1998 also showed that his galaxy and circumnuclear starburst data
could be fit by a star formation law with direct dependence on
galactic orbital dynamics:
\begin{equation}
\label{sfromega}
\Sigma_{\rm sfr} =  B_{\rm \Omega} \Sigma_g \Omega,
\end{equation}
hereafter the Gas-$\Omega$ law, where $B_{\Omega}=0.017$ and $\Omega$
is evaluated at the outer radius that is used to perform the disk
averages. This law has also been studied in samples of galaxies and
starbursts extending to higher redshifts by, for example, Genzel et
al. (2010), Daddi et al. (2010) and Garc\'ia-Burillo et al. (2012). It
implies that a fixed fraction, about 10\%, of the gas is turned into
stars every outer orbital timescale of the star-forming disk and
motivates theoretical models that relate star formation activity to
the dynamics of galactic disks.


In order to link global galactic dynamics with the scales of
star-forming regions in GMCs, Tan (2000) proposed a model of star
formation triggered by GMC collisions in a shearing disk, which
reproduces eq. (\ref{sfromega}) in the limit of a flat rotation curve
since the collision time is estimated to be a short and approximately
constant fraction, $\sim 20\%$, of the orbital time, $t_{\rm
  orbit}$. This behavior of the GMC collision time was confirmed in
the numerical simulations of Tasker \& Tan (2009). The GMC Collision
model assumes a Toomre $Q$ parameter of order unity in the
star-forming part of the disk, a significant fraction (e.g. $\gtrsim
1/3$) of total gas in gravitationally bound clouds, and a velocity
dispersion of these clouds set by gravitational scattering (Gammie et
al. 1991). Then, the predicted SFR is
\begin{equation}
\label{sfrcoll}
\Sigma_{\rm sfr} = B_{\rm CC} Q^{-1} \Sigma_g \Omega (1 - 0.7 \beta), \:\:\: (\beta \ll 1)
\end{equation}
where $\beta\equiv d\:{\rm ln}\:v_{\rm circ} / d\:{\rm ln}\: r$ and
$v_{\rm circ}$ is the circular velocity at a particular galactocentric
radius $r$. Note that $\beta=0$ for a flat rotation curve. There is a
prediction of reduced star formation efficiencies per orbit compared
to eq. (5) in regions with reduced shear, i.e. typically the inner
parts of disk galaxies.

The above six laws are not the only ones that have been proposed. For
example, Ostriker, McKee, and Leroy (2010) suggested a self-regulated
star formation model for disk galaxies (though mostly focussed on the
more \ion{H}{1}-rich outer regions compared to our present study),
$\Sigma_{\rm sfr} \propto \Sigma_g \sqrt{\rho_{\rm sd}}$, where
$\rho_{\rm sd}$ is the midplane volume density of stars and dark
matter.  However, this volume density is difficult to evaluate
empirically.

Paper I showed that the KMT2009 turbulence-regulated, constant
molecular, and GMC collision models can produce the observed SFRs with
rms error of about a factor of 1.5, where each galaxy is allowed one
free parameter. The other models do moderately worse with larger rms
error of factor of 1.8 and 2.0 for Gas-$\Omega$ and KM2005
turbulence-regulated models, respectively.

\begin{deluxetable}{ccccccccc}
\tabletypesize{\footnotesize}
\tablecolumns{9}
\tablewidth{0pt}
\tablecaption{Basic properties of sample galaxies}
\tablehead{\colhead{Galaxy} &
           \colhead{Messier} &
           \colhead{Morphological} &
           \colhead{$d$} &
           \colhead{$r_{\rm B25}$} &
           \colhead{$r_{\rm out}$} &
           \colhead{$r_{\rm ext}$} &
           \colhead{$\bar{v}_{\rm circ}$} &
           \colhead{$\bar{\beta}$}\\
	   \colhead{NGC:} &
	   \colhead{Index} &
	   \colhead{Type} &
	   \colhead{(Mpc)} &
	   \colhead{(kpc)} &
	   \colhead{(kpc)} &
	   \colhead{(kpc)} &
	   \colhead{(${\rm km\: s^{-1}}$)} &
           \colhead{}}
\startdata
0628$^*$ & M74 & SA(s)c & 9.7 & 10.4 & 4.5 & 5.6 & 181 & 0.272\\
2841$^*$ & - & SA(r)b & 15.4 & 14.2 & 7.8 & 10.0 & 300 & 0.022\\
2903 & - & SAB(rs)bc & 9.3 & 15.2 & 5.2 & 5.7 & 145 & 0.410\\
3184$^*$ & - & SAB(rs)cd & 12.6 & - & 5.4 & 6.8 & 119 & 0.608\\
3351$^*$ & M95 & SB(r)b & 9.9 & 10.6 & 5.6 & 7.5 & 171 & 0.209\\
3521$^*$ & - & SAB(rs)bc & 11.5 & 12.9 & 5.9 & 6.9 & 175 & 0.367\\
3627$^*$ & M66 & SAB(s)b & 10.1 & 13.8 & 8.1 & 8.4 & 164 & 0.237\\
4254 & M99 & SA(s)c & 15.6 & 14.6 & 9.1 & 12.3 & 150 & 0.337\\
4321 & M100 & SAB(s)bc & 15.8 & 12.5 & 8.8 & 11.0 & 176 & 0.320\\
4579 & M58 & SAB(rs)b & 19.4 & 15.0 & 10.7 & 11.4 & 209 & 0.320\\
4736$^*$ & M94 & (R)SA(r)ab & 4.9 & 5.3 & 2.0 & 4.2 & 145 & 0.118\\
5055$^*$ & M63 & SA(rs)bc & 8.2 & 17.3 & 8.4 & 10.0 & 177 & 0.132\\
5194$^*$ & M51 & SA(s)bc pec & 8.1 & 9.0 & 6.9 & 7.9 & 195 & 0.181\\
5457 & M101 & SAB(rs)cd & 7.0 & 25.8 & 6.9 & 9.0 & 182 & 0.270\\
6946$^*$ & - & SAB(rs)cd & 5.5 & 9.8 & 6.0 & 7.5 & 145 & 0.351\\
7331$^*$ & - & SA(s)b & 14.5 & 19.5 & 7.2 & 9.0 & 202 & 0.284\\
\enddata

\tablenotetext{*}{Analyzed in Paper I}

\label{tb:1}
\end{deluxetable}

\section{Methodology}\label{S:method}

We utilize data presented by Leroy et al. (2013), providing
$\Sigma_{\rm H2}$, $\Sigma_{\rm HI}$, $\Sigma_{\rm sfr}$, $v_{\rm
  circ}$, and $\beta$ of 30 nearby disk galaxies 
(see also Schruba et al. 2011).
We refer the reader to Leroy et al. (2013) for the methods used to
estimate these quantities. We note that $v_{\rm circ}$ and $\beta$ are
based on simple parameterized fits to tilted ring modeling (de Blok et
al. 2008) based on HI (Walter et al. 2008) and CO data. The fits wash
out $\sim$kpc-scale variations in the rotation curve.

We focus only on the galaxies which have some regions
where molecular gas dominates over atomic, $\Sigma_{\rm H2}\ge
\Sigma_{\rm HI}$. There are 21 galaxies
that fulfill this criterion, including the 12 galaxies analyzed in
Paper I. Our focus is on ``large'' disk galaxies, so we restrict
further analysis to systems with $\bar{v}_{\rm circ}> 100\:{\rm km
  s}^{-1}$ in the molecular dominated region.  One motivation for this
is to exclude dwarf galaxies, which may have larger systematic
differences in properties such as metallicity that can affect
estimates of molecular gas mass.
This leaves 16 galaxies in our sample: 5 new galaxies compared to
Paper I, and 1 galaxy from Paper I (NGC 3198) now excluded.

Table 1 lists the basic properties of our galaxy sample.
The morphological types and distances are assessed from the NASA/IPAC
Extragalactic Database (NED). Galactic radii ($r_{\rm B25}$) are
adopted from Leroy et al. (2013). Finally, the last two columns show
average circular velocity from the rotation curve, $\bar{v}$, and
average logarithmic derivative of the rotation curve, $\bar{\beta}$.
Typically, galaxies in our sample have $\bar{\beta}$ $\lsim$ 0.4 and
$\bar{v}$ $\gsim$ 150 km $s^{\rm -1}$ except NGC 3184 that has
$\bar{\beta}$ $\sim$ 0.6 and $\bar{v}_{\rm circ}$ $\sim 120\:{\rm
  km\:s^{-1}}$.


Following the method of Paper I, we fit the observed data of molecular
dominated regions of the sample galaxies with the six star formation
laws described in \S\ref{S:intro} to derive the best-fit values of
A$_{\rm g}$, A$_{\rm H2}$, A$_{\rm KM}$, A$_{\rm KMT}$, B$_{\rm
  \Omega}$, and B$_{\rm CC}$, from a Kennicutt-Schmidt law
(Eq.~\ref{sfr1}), using $\alpha_{\rm g}$ = 1.5, the Constant Molecular
law (Eq.~\ref{sfrH2}), the KM05 turbulence regulated law
(Eq.~\ref{sfrKM}), calculating the orbital angular frequency,
$\Omega$, from the given $v_{\rm circ}/r$ and setting the value of
$Q=1$, a KMT09 turbulence regulated law (Eq.~\ref{sfrKMT}), the
Gas-$\Omega$ law (Eq.~\ref{sfromega}), and the GMC Collision law
(Eq.~\ref{sfrcoll}), setting $Q =1.5$. The outer radius, $r_{\rm
  out}$, of the sample galaxies is determined by the radius where
molecular gas dominates over atomic gas, $\Sigma_{\rm H2} \ge
\Sigma_{\rm HI}$. For NGC 2814, molecular gas is not dominant over
atomic gas in the central region but it becomes so at about 2~kpc.

The best-fit star formation law parameters are constrained by
comparing $\Sigma_{\rm sfr, theory}$ from these six star formation
laws with the observed $\Sigma_{\rm sfr, obs}$. For each galaxy we
derive $\chi$, where $\chi^2 \equiv (N_{\rm ann} - N_{\rm fit})^{-1}
\sum({\rm log}_{10}\: R_{\rm sfr})^2$, $N_{\rm ann}$ is the number of
resolved annuli in the galaxy, $R_{\rm sfr} =
\Sigma_{\rm sfr,theory}/\Sigma_{\rm sfr,obs}$, and $N_{\rm fit}=1$
(note, each star formation law has one free parameter). We also carry
out this analysis for the entire sample and for sub-samples in two
ways. First each galaxy is allowed one free parameter so that $N_{\rm
  fit}$ equals the number of galaxies in the sample (i.e., 16) or
sub-sample. Second, we fit for a single star formation law for the
sample or sub-sample with one global free parameter ($N_{\rm
  fit}=1$). The values of $\chi$ and the rms dispersions of the data
about the best fits are considered.

We repeat the above analysis for ``molecular rich'' regions with
$\Sigma_{\rm HI}/2<\Sigma_{\rm H2}<\Sigma_{\rm HI}$, which typically
applies to an extended annulus in the galaxy out to a radius $r_{\rm
  ext}$. Only galaxies with $N_{\rm ann}\geq 3$ in the molecular rich
region are analyzed, i.e. 14 galaxies. Finally, we also carry out the analysis
on the combined molecular dominated and rich regions for all the galaxies.

With 16 galaxies, we are now in a position to also examine trends of
star formation law parameters with galaxy properties, both by defining
and comparing sub-samples and looking for correlations with continuous
variables.


\newpage
\section{Results}\label{S:results}

\subsection{Test of Star Formation Laws}\label{S:lawtest}

\begin{figure}
\centering
\begin{tabular}{cc}
\epsfig{file=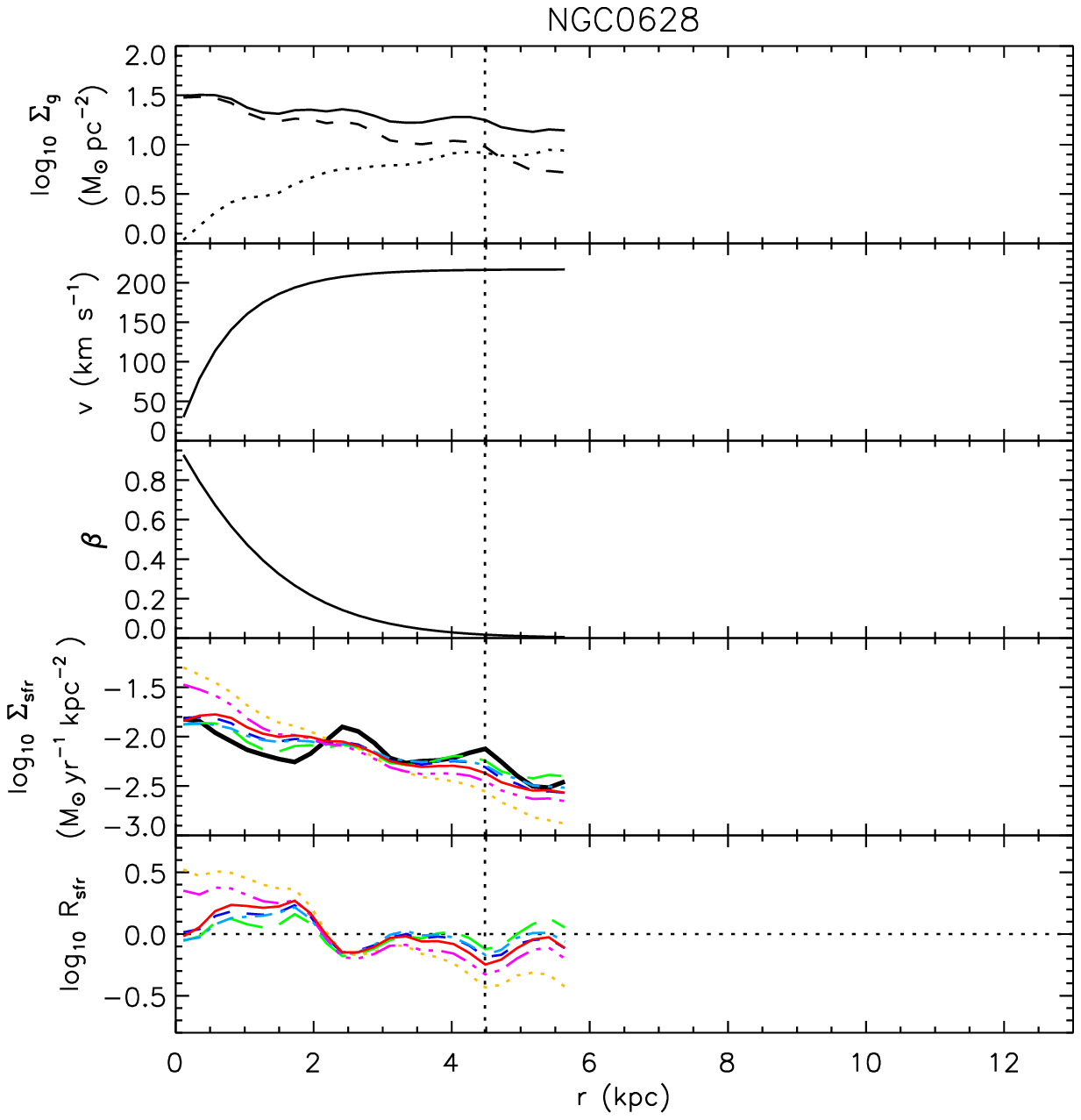,width=0.45\linewidth,clip=} & 
\epsfig{file=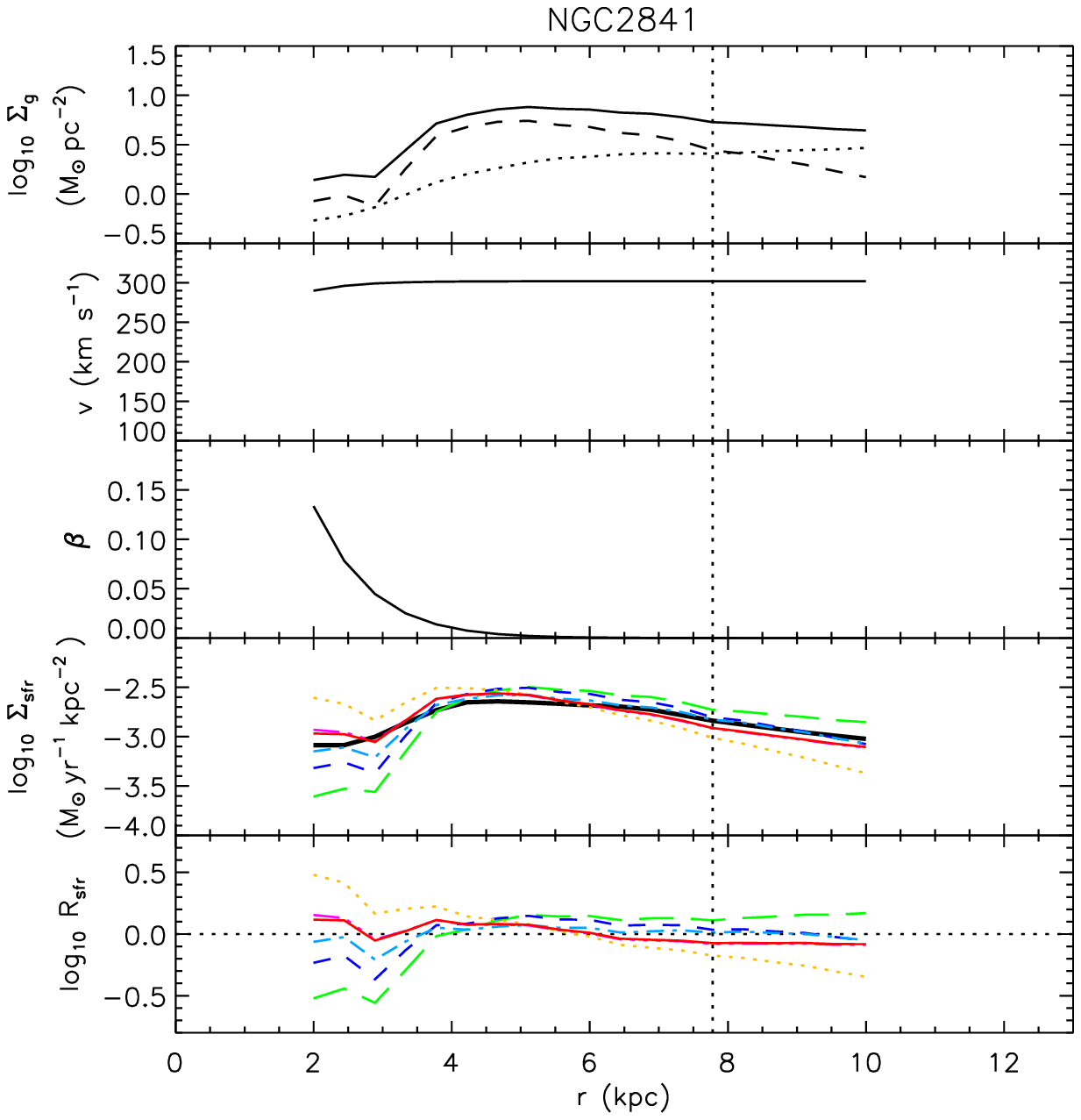,width=0.45\linewidth,clip=} \\
\epsfig{file=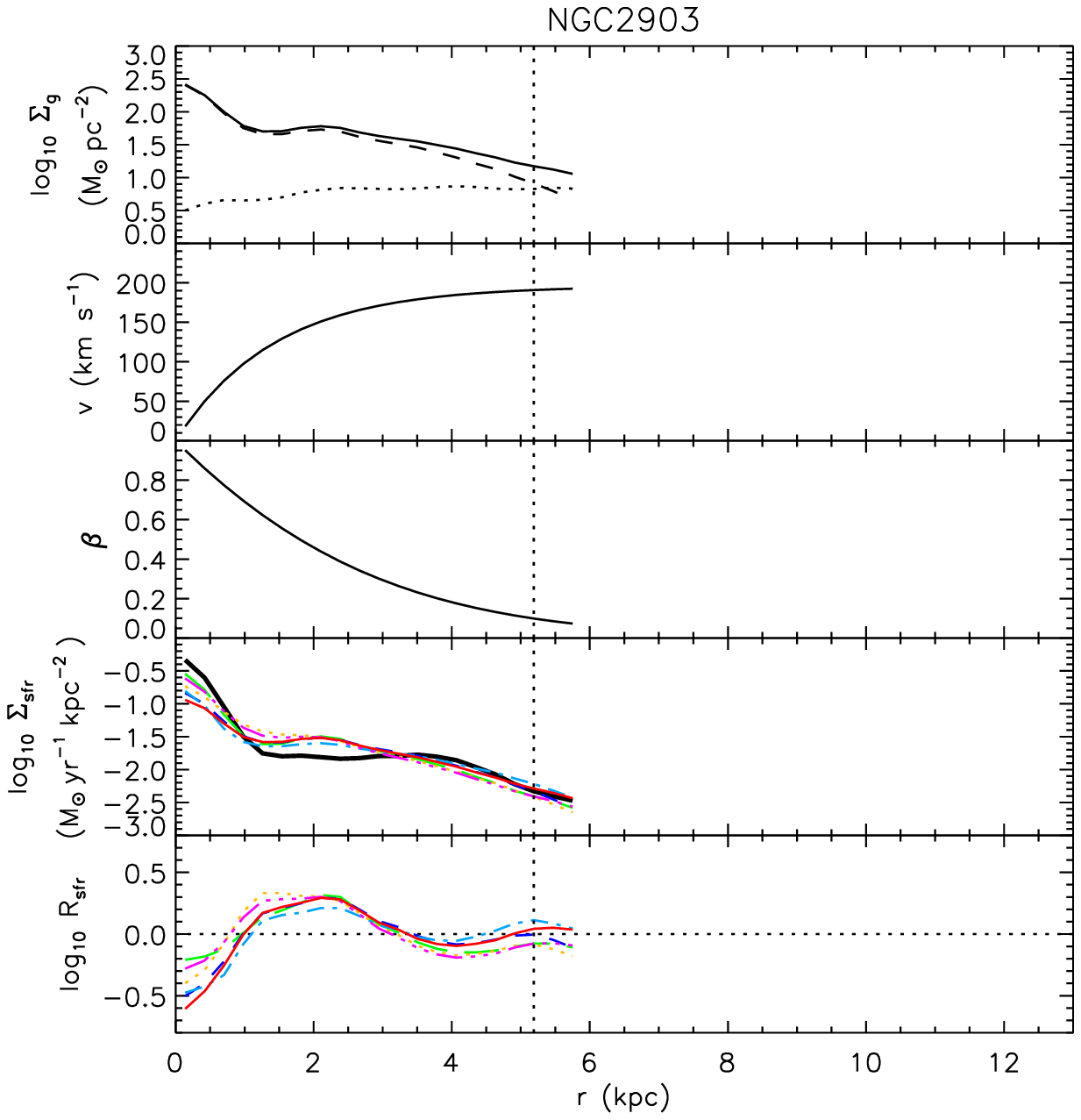,width=0.45\linewidth,clip=} &
\epsfig{file=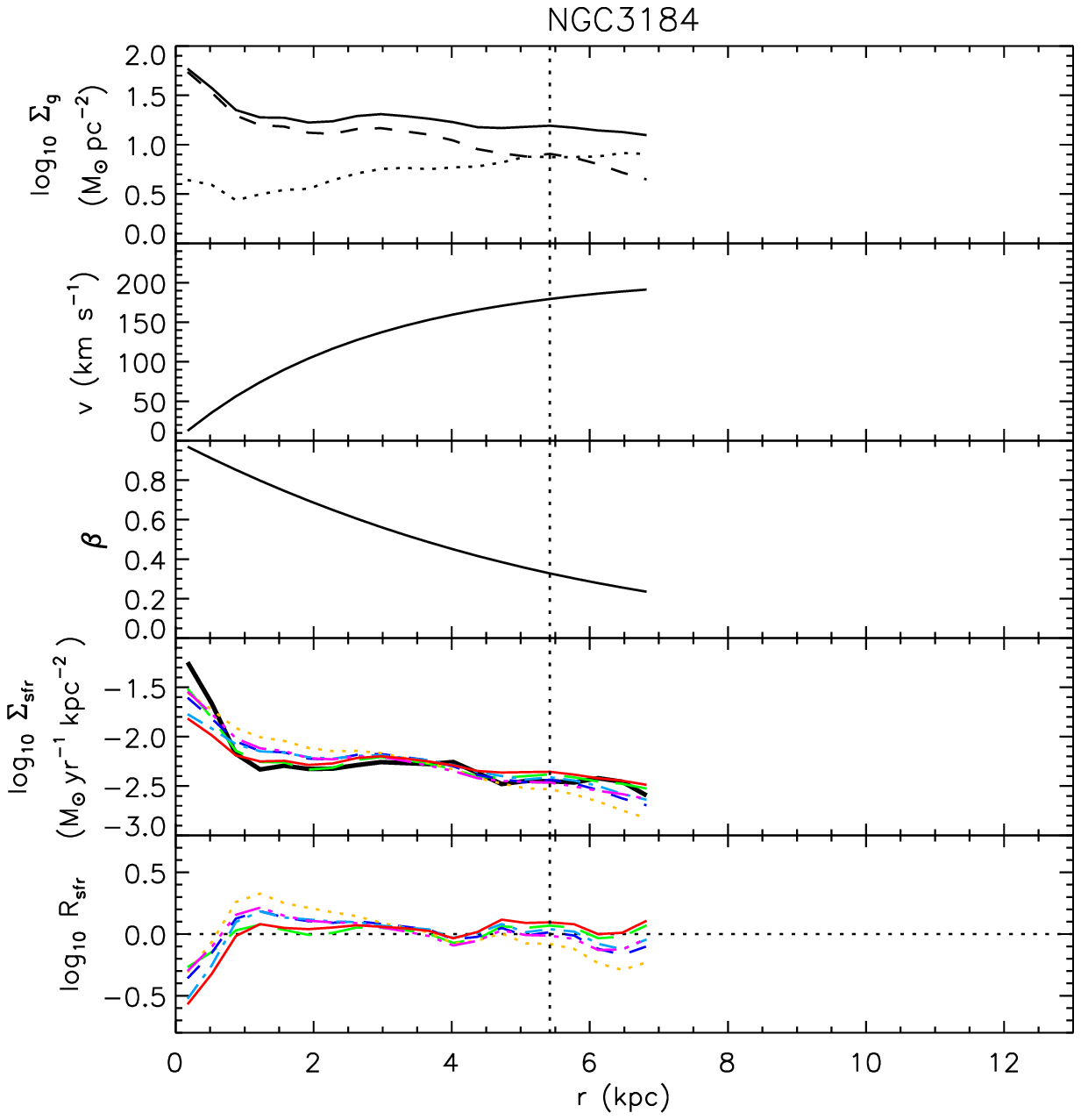,width=0.45\linewidth,clip=}
\end{tabular}

\caption{
Radial distribution of properties of NGC 0628, NGC 2841, NGC 2903, and
NGC 3184, as indicated, for the regions where $\Sigma_{\rm H2} \ge
\Sigma_{\rm HI}/2 $. In each 5-panel figure, the top panel shows
radial profiles of surface density of molecular hydrogen, $\Sigma_{\rm
  H2}$, (dashed), atomic hydrogen, $\Sigma_{\rm HI}$, (dotted) and
total gas (solid). The dotted vertical line indicates $r_{\rm out}$,
where $\Sigma_{\rm H2}$ becomes less than $\Sigma_{\rm HI}$. The
second and third panels show the rotation velocity curve, $v$, and its
logarithmic derivative, $\beta$, respectively. The fourth panel shows
the predicted SFR surface density compared with the observed data
(thick-dotted). Each star formation law is represented by:
Kennicutt-Schmidt (Eq.~1; green long-dashed), Constant Molecular
(Eq.~2; blue dashed), KM05 (Eq.~3; orange dotted), KMT09 (Eq.~4; cyan
dot-dashed), Gas-$\Omega$ (Eq. 5; magenta dot-dot-dashed), and GMC
Collision law (Eq. 6; red solid). Finally, the fifth panel shows ${\rm
  log}_{10}\:R_{\rm sfr}$, i.e., ${\rm log}_{10}$ of the ratio of the
predicted SFR surface densities to the observed surface densities.
}\label{fig:1}
\end{figure}

\begin{figure}
\centering
\begin{tabular}{cc}
\epsfig{file=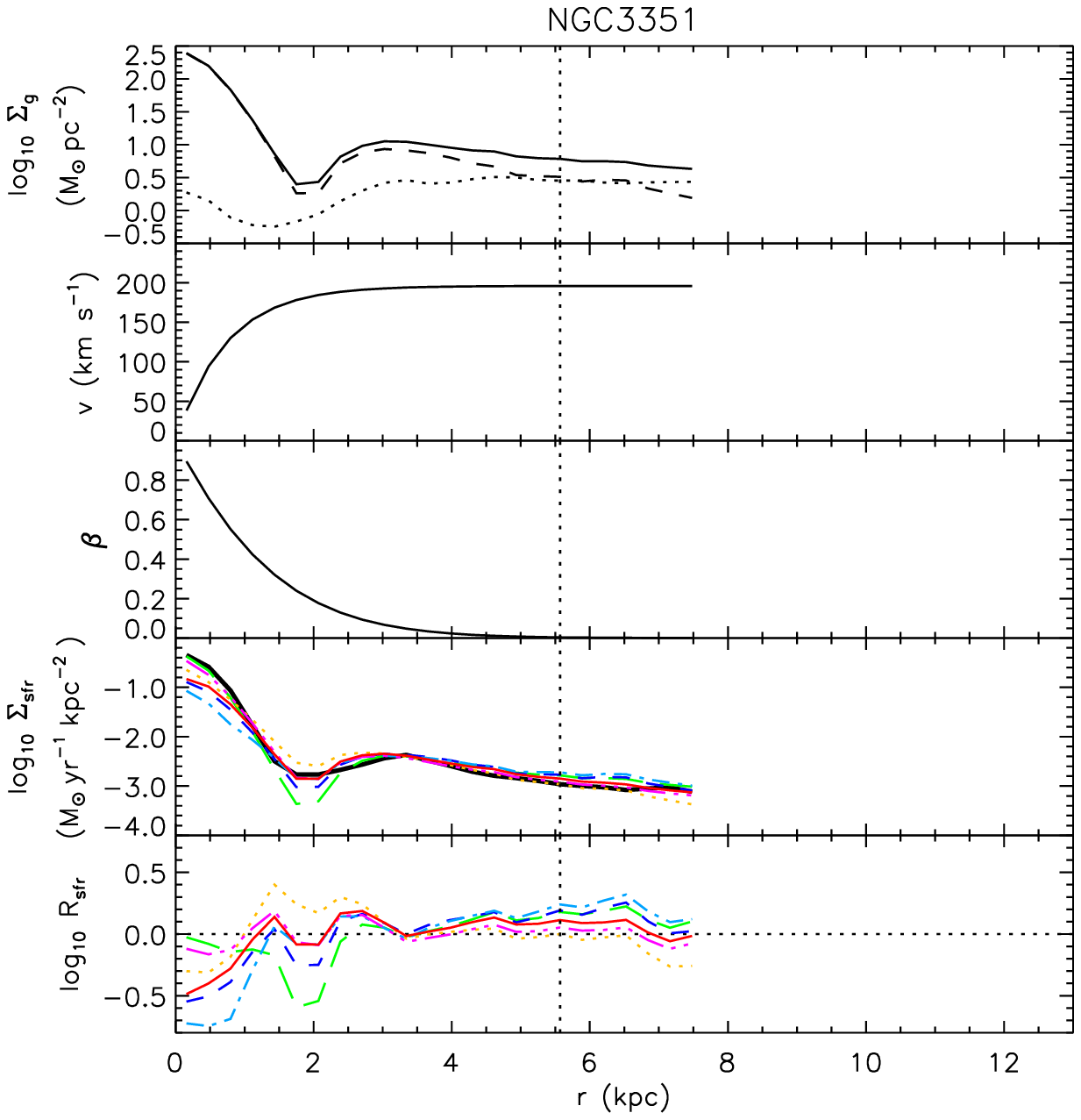,width=0.45\linewidth,clip=} & 
\epsfig{file=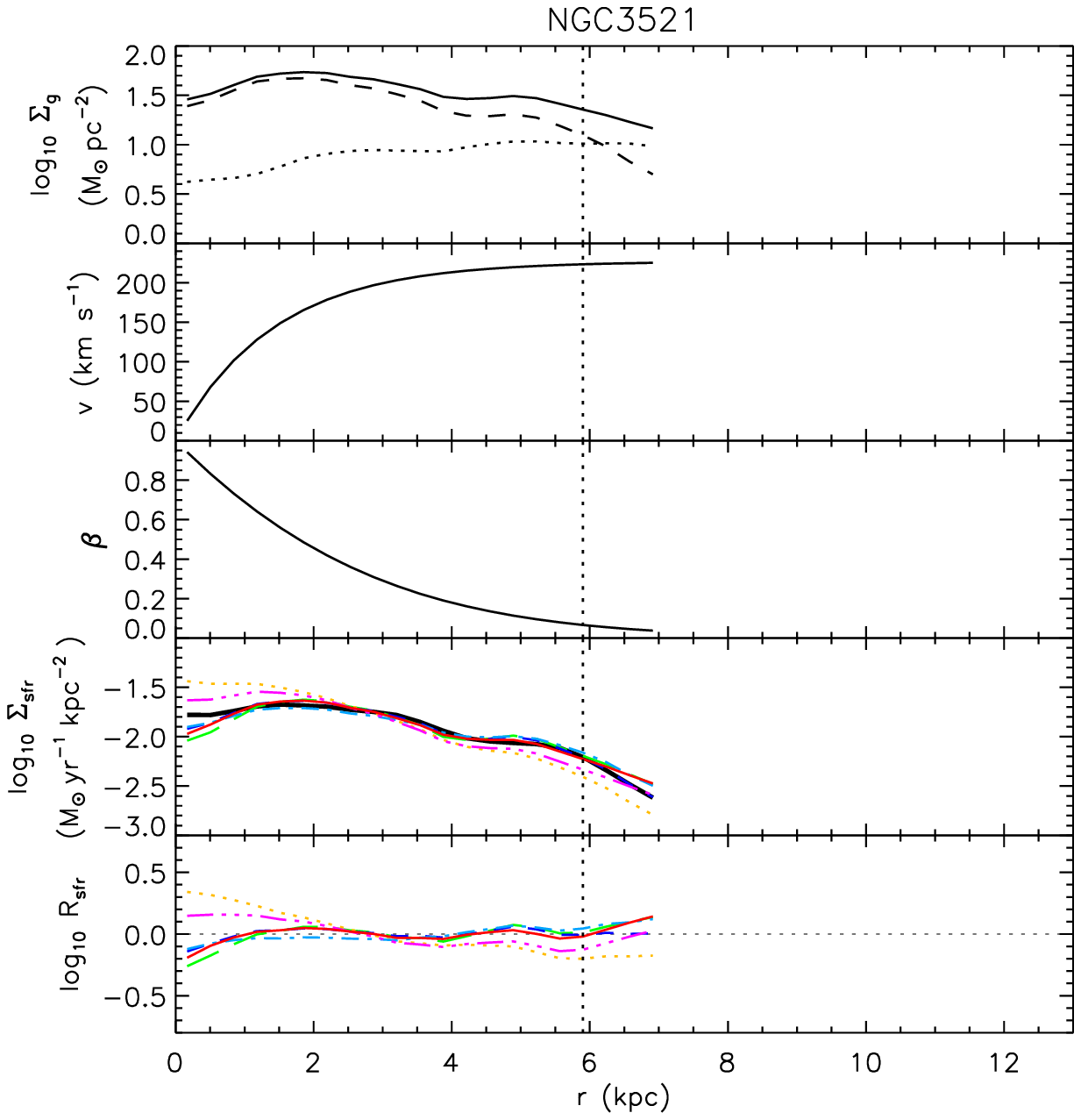,width=0.45\linewidth,clip=} \\
\epsfig{file=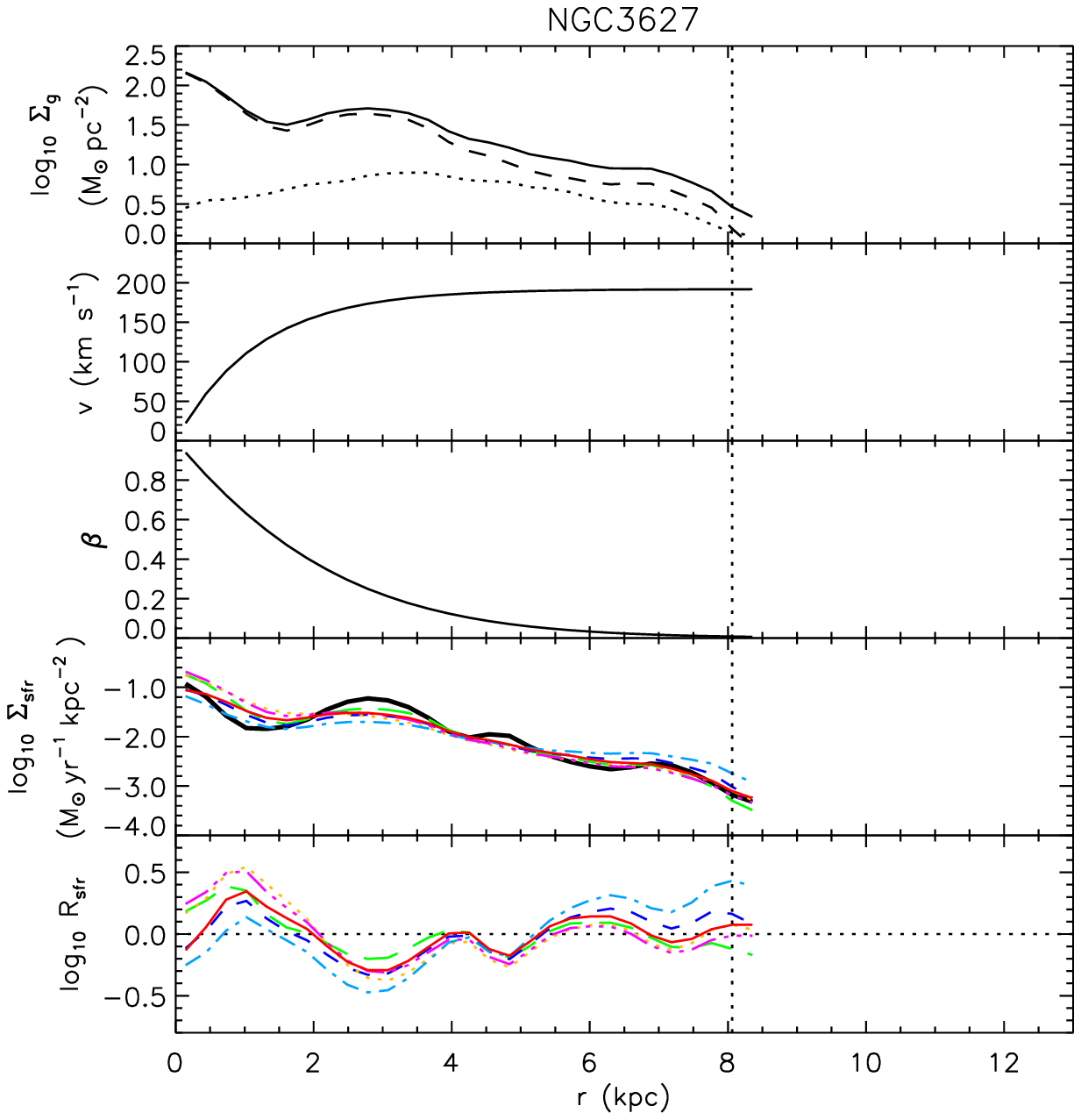,width=0.45\linewidth,clip=} &
\epsfig{file=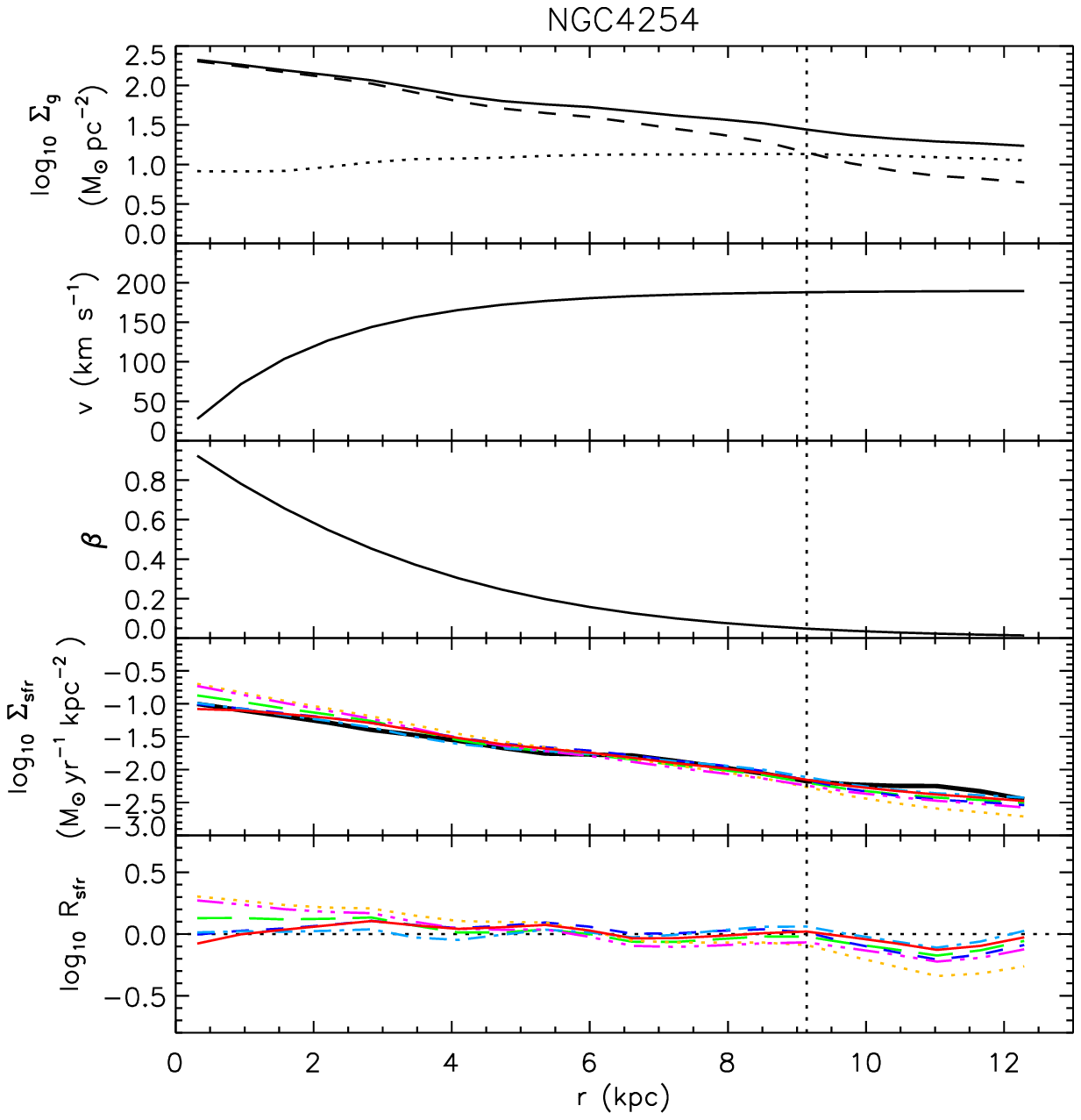,width=0.45\linewidth,clip=}
\end{tabular}
\caption{
Radial distribution of properties of NGC 3351, NGC 3521, NGC 3627, and
NGC 4254, with labeling as in Figure~\ref{fig:1}.}\label{fig:2}
\end{figure}

\begin{figure}
\centering
\begin{tabular}{cc}
\epsfig{file=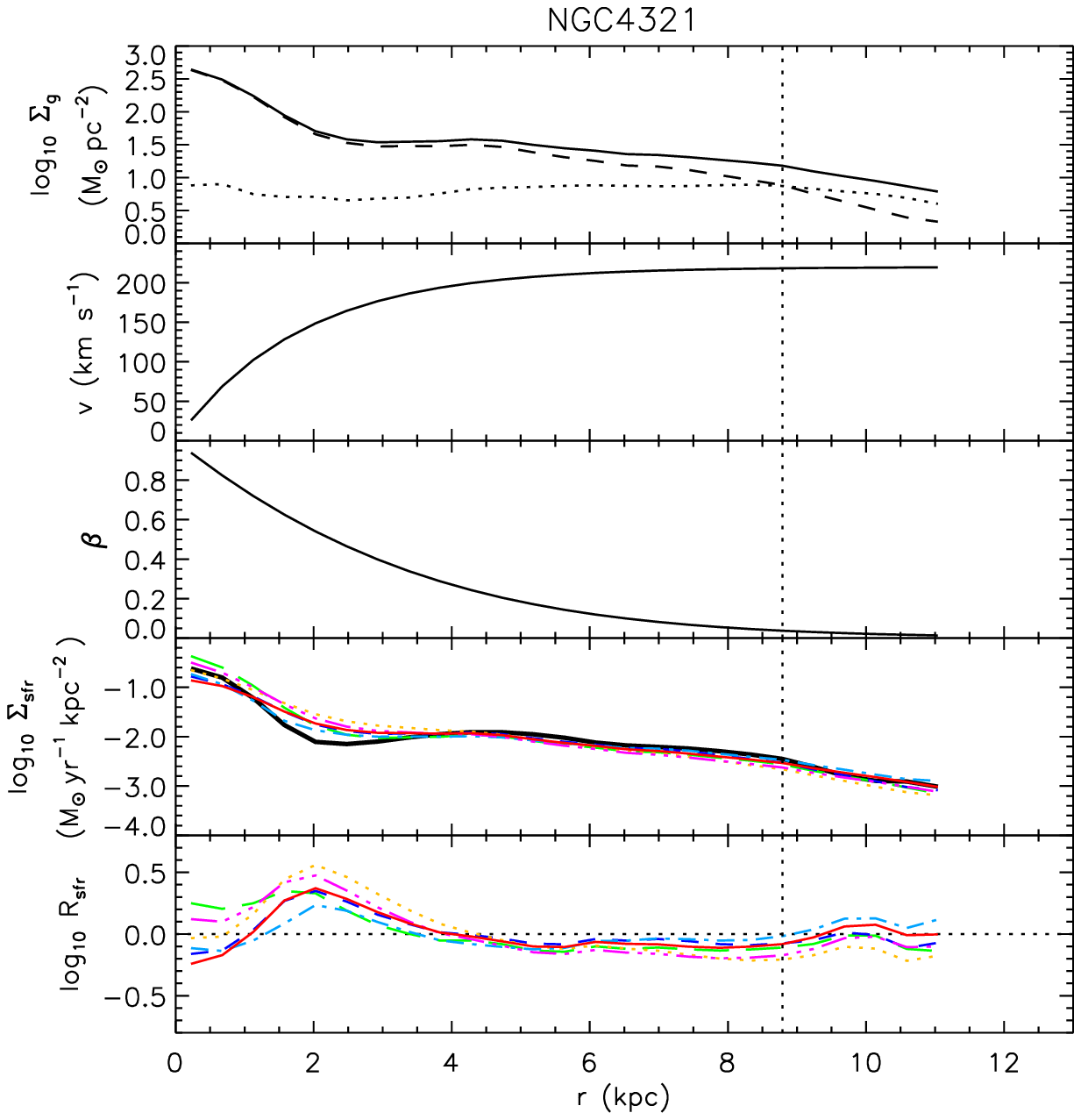,width=0.45\linewidth,clip=} & 
\epsfig{file=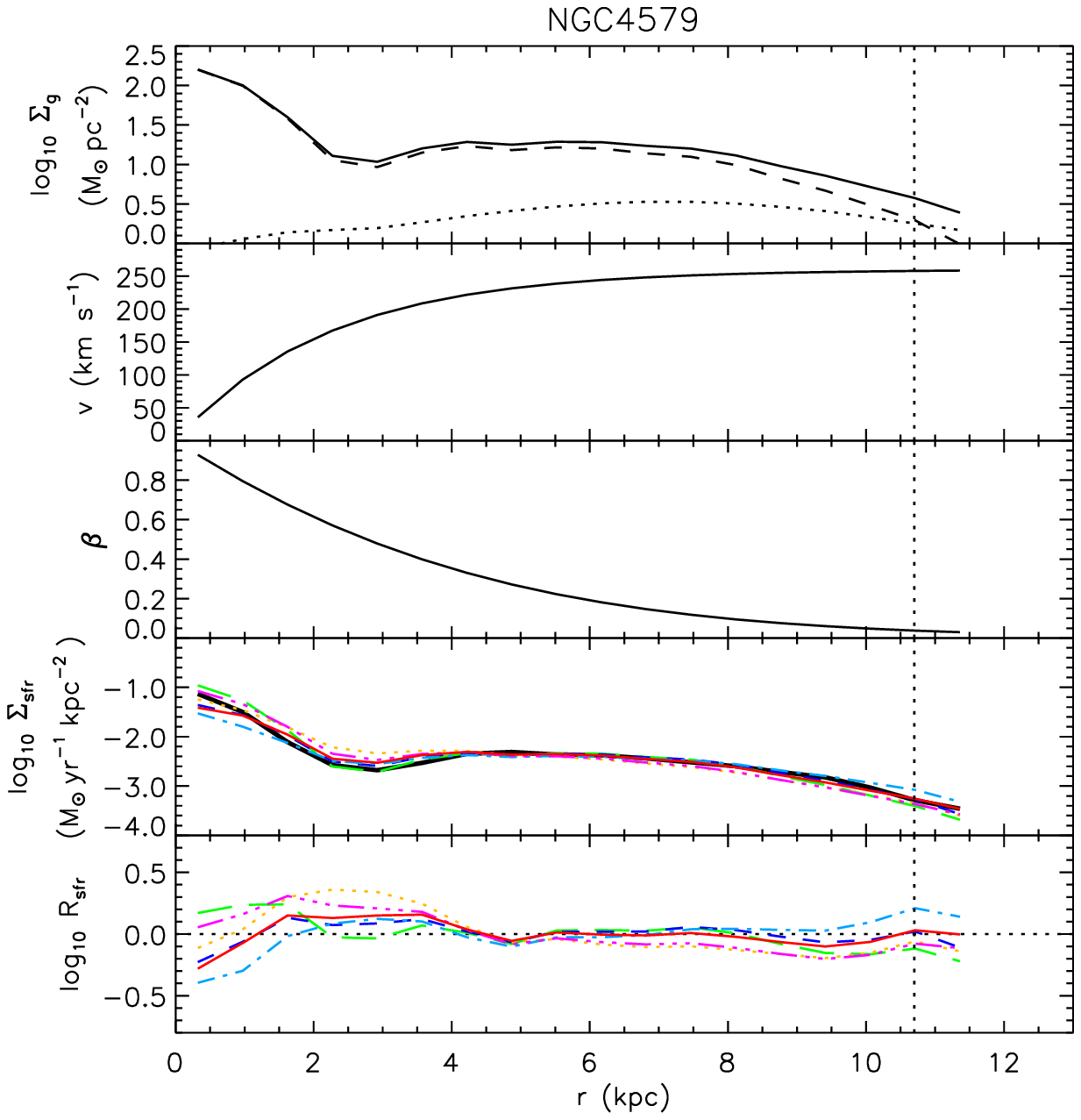,width=0.45\linewidth,clip=} \\
\epsfig{file=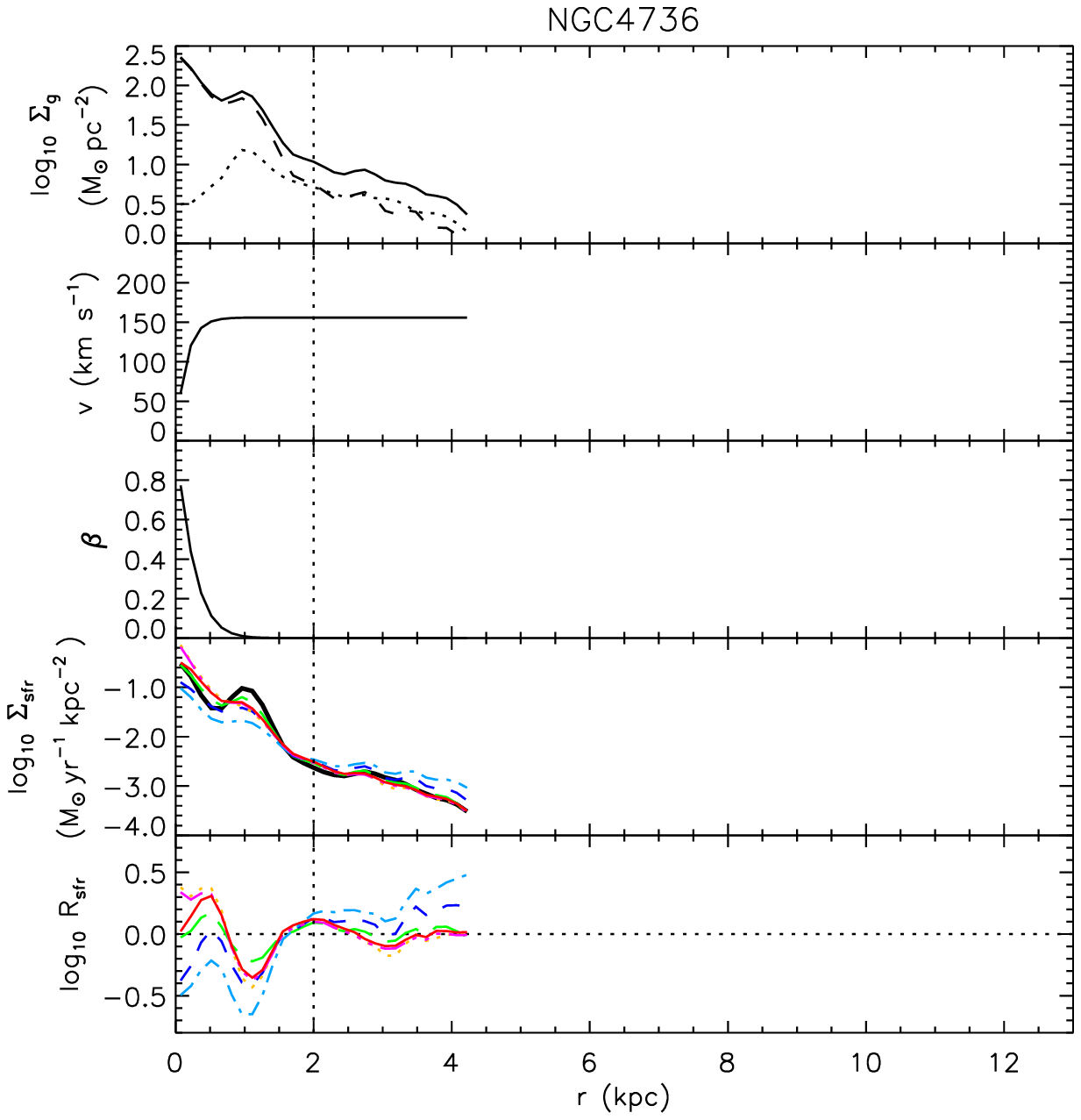,width=0.45\linewidth,clip=} &
\epsfig{file=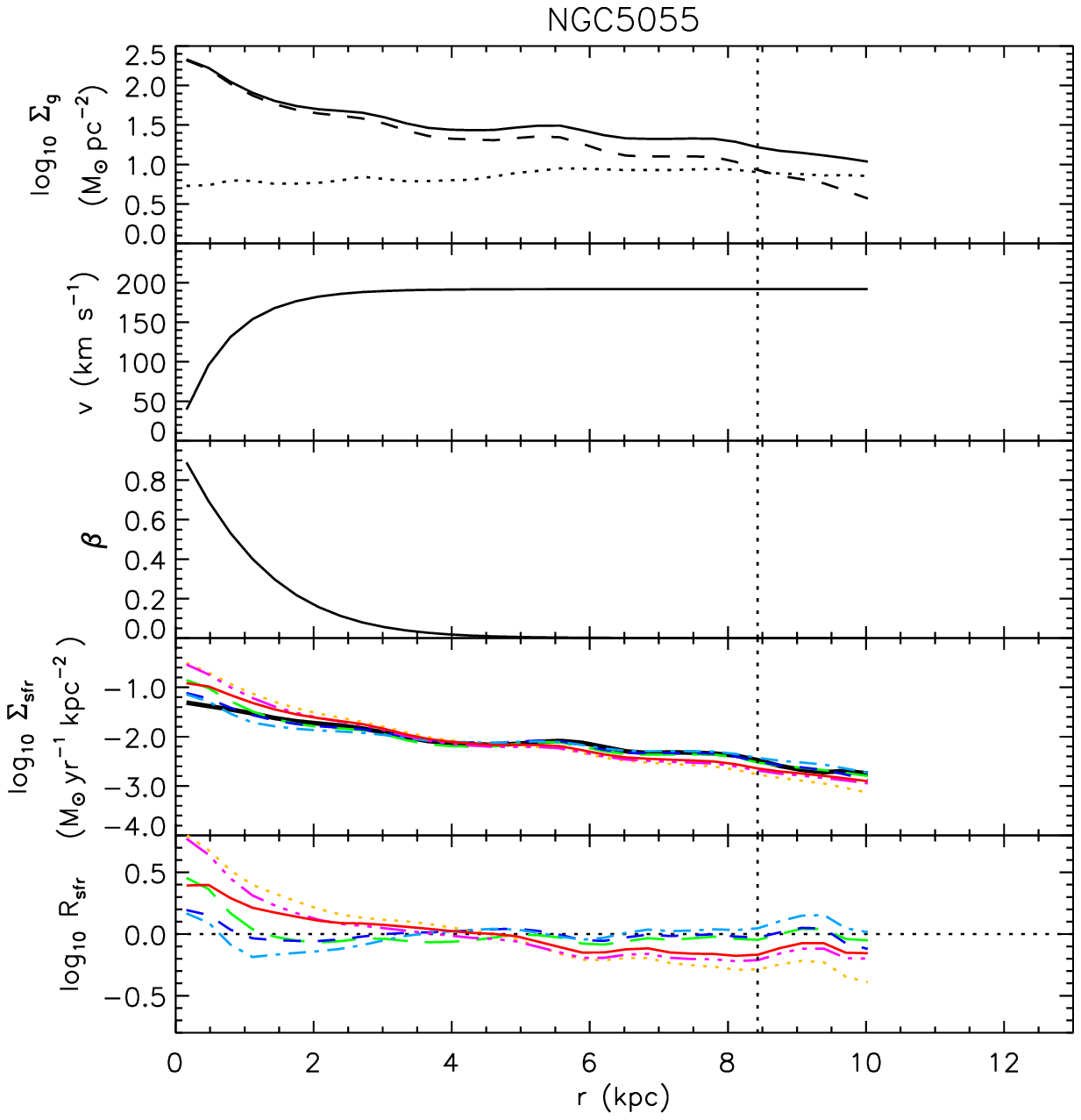,width=0.45\linewidth,clip=}
\end{tabular}
\caption{
Radial distribution of properties of NGC 4321, NGC 4579, NGC 4736, and
NGC 5055, with labeling as in Figure~\ref{fig:1}.}\label{fig:3}
\end{figure}

\begin{figure}
\centering
\begin{tabular}{cc}
\epsfig{file=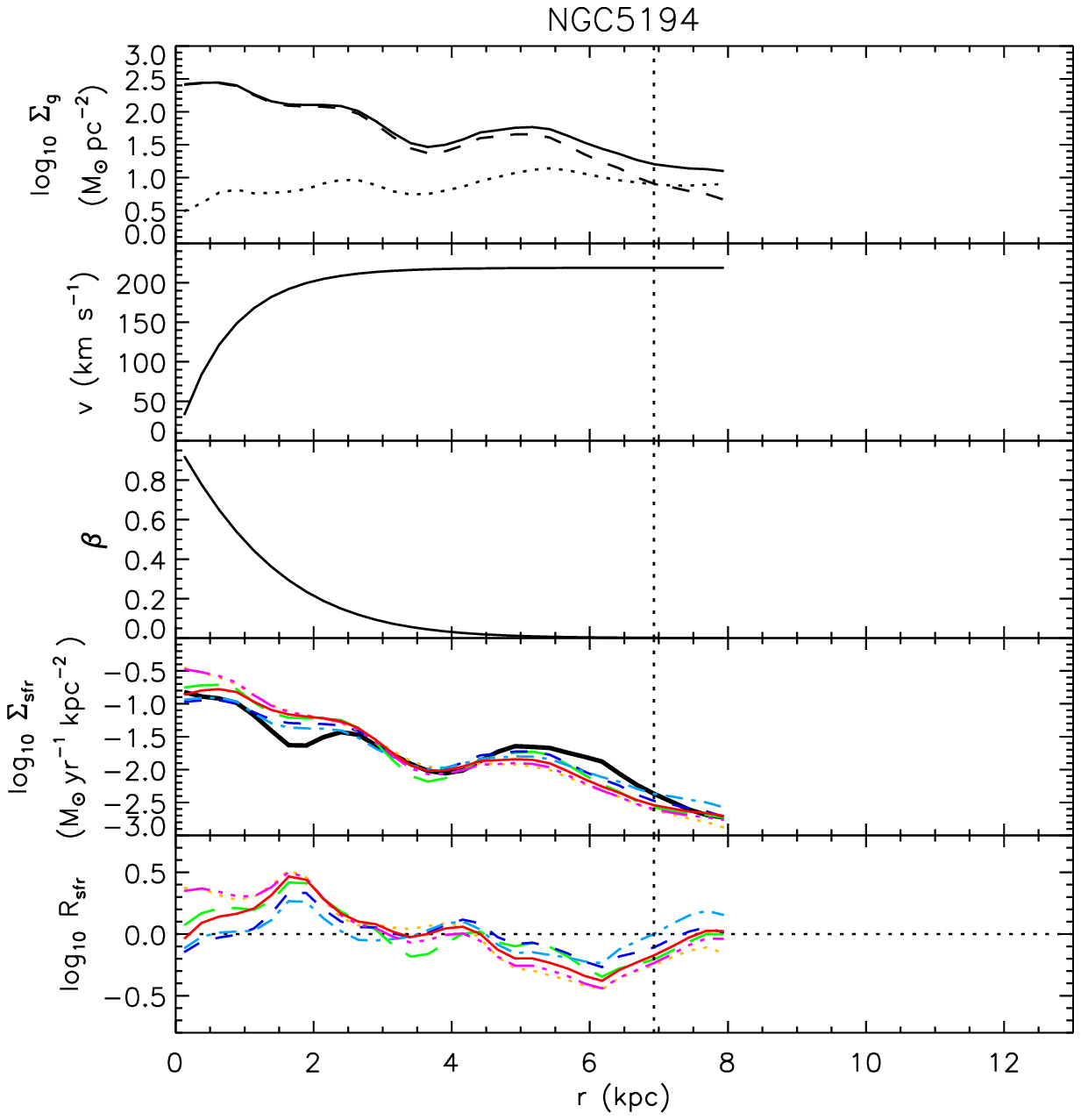,width=0.45\linewidth,clip=} & 
\epsfig{file=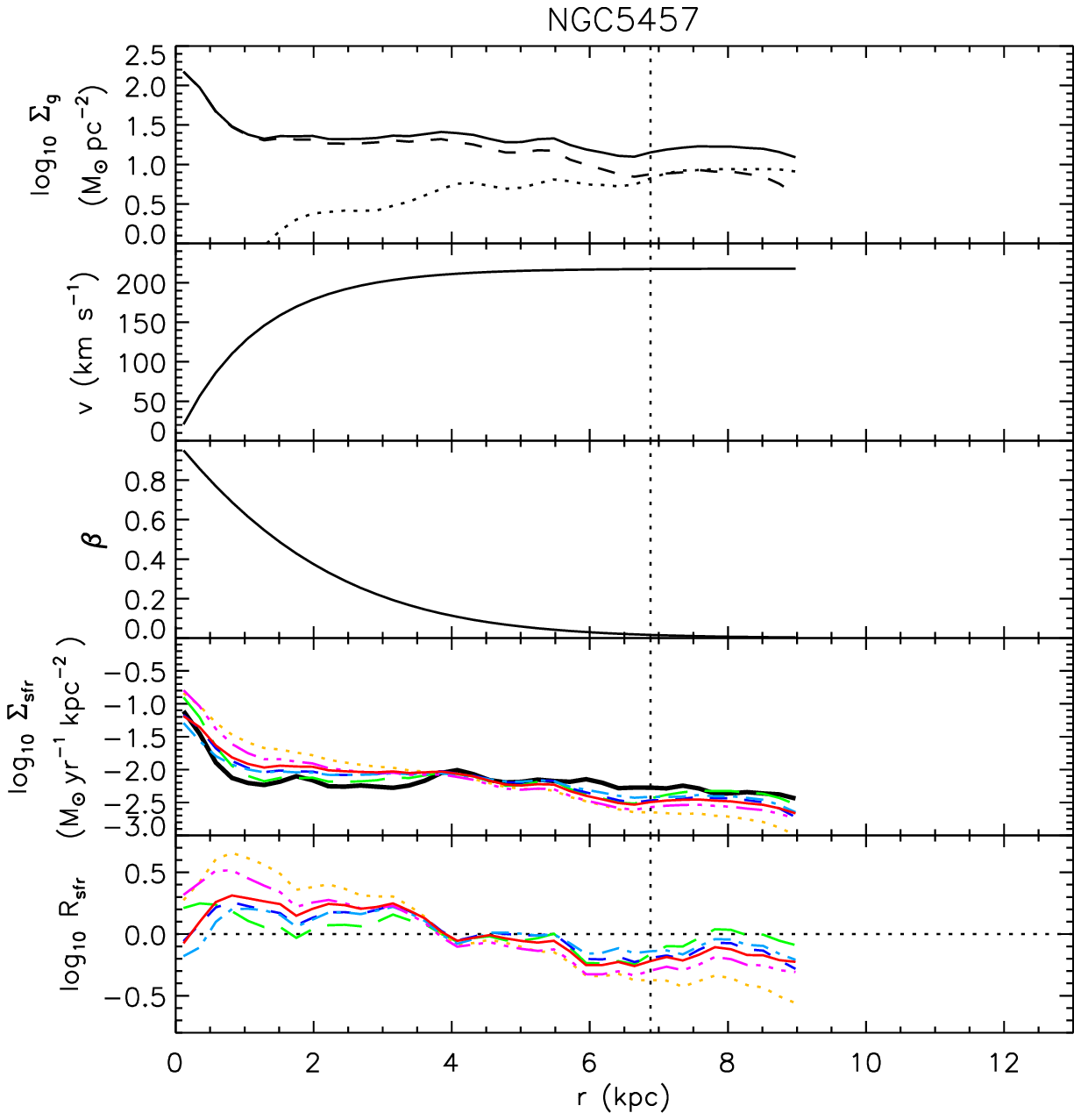,width=0.45\linewidth,clip=} \\
\epsfig{file=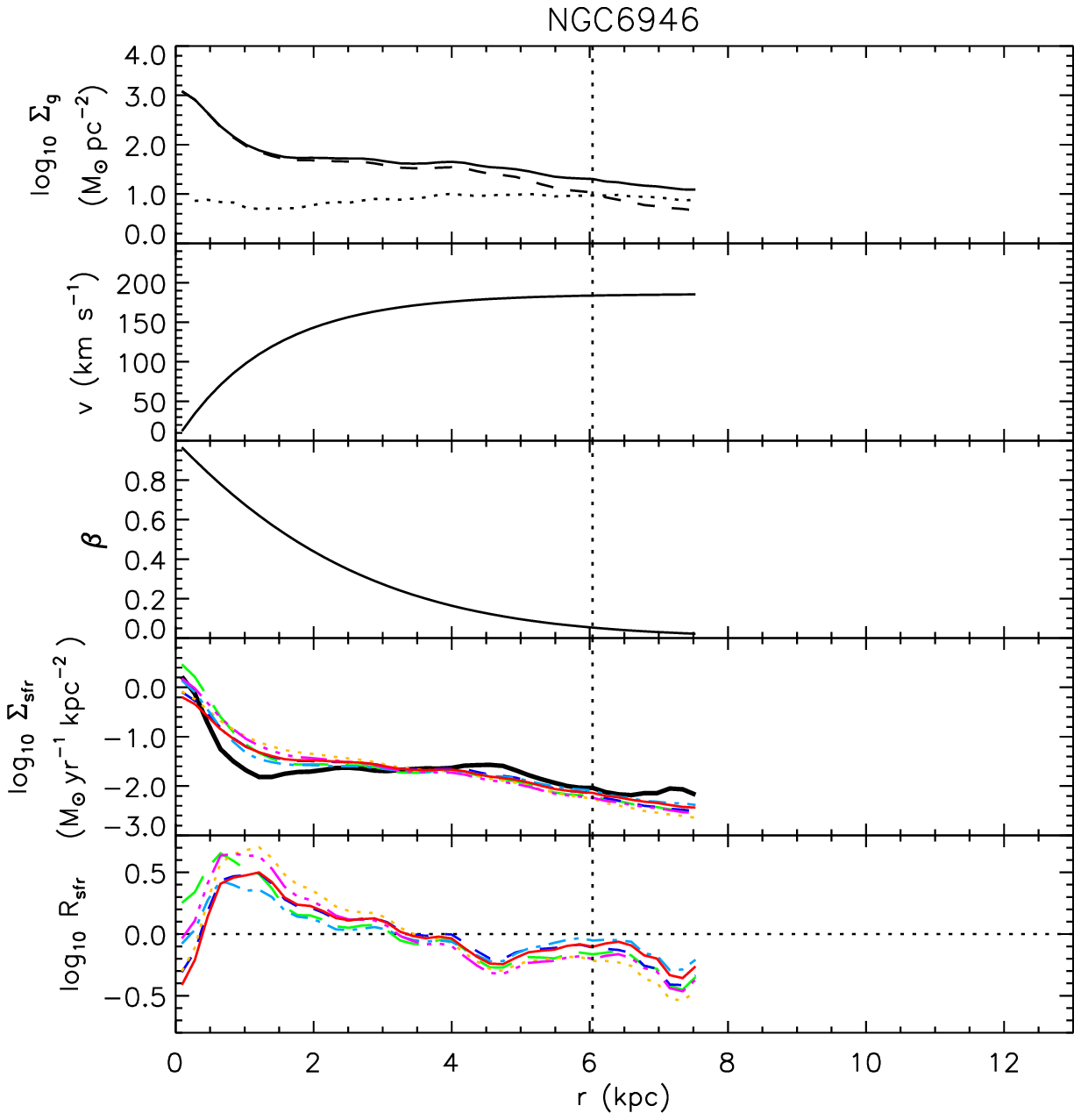,width=0.45\linewidth,clip=} &
\epsfig{file=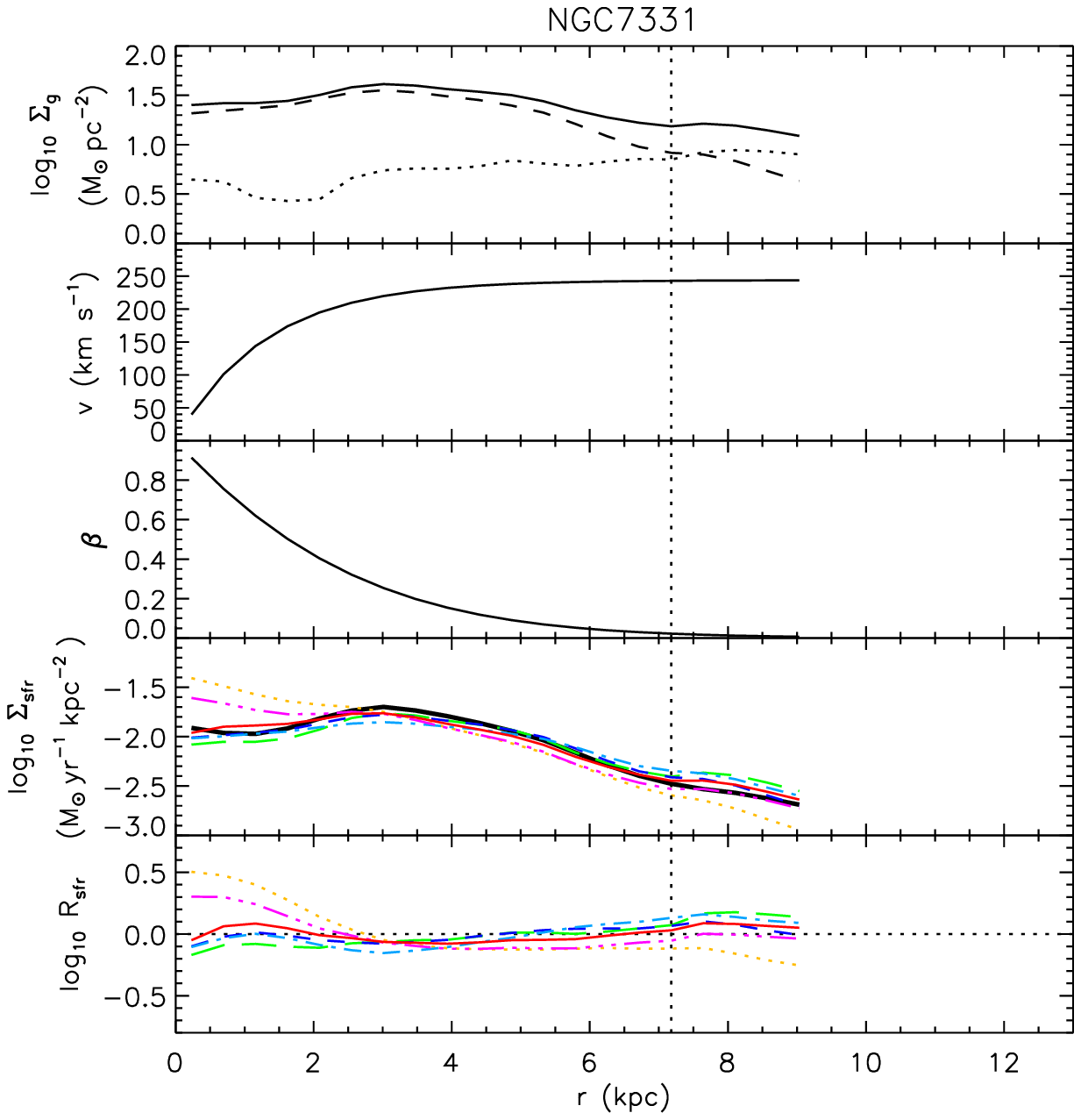,width=0.45\linewidth,clip=}
\end{tabular}
\caption{
Radial distribution of properties of NGC 5194, NGC 5457, NGC 6946, and
NGC 7331, with labeling as in Figure~\ref{fig:1}.}\label{fig:4}
\end{figure}

First, we examine how well the star formation laws described in
\S\ref{S:laws} do in predicting the SFR as a function of
galactocentric radius, given their required inputs. Note that these
input requirements differ: e.g. the Constant Molecular law only needs
the surface density of molecular gas, while some of the other laws
require multiple inputs, each of which has inherent observational
uncertainties. Thus, while the relative accuracy with which the laws
can predict SFRs is still interesting (e.g., if one is concerned with
how accurate the use of given law will be in a model of galaxy
evolution), this relative ordering may not necessarily distinguish
between which physical mechanism(s) is responsible for setting SFRs. In
addition, there are different levels of physics built into these
laws. The Constant Molecular law uses an input that is relatively
close to star formation, namely the amount of molecular gas, without
trying to predict why certain regions have a given molecular
content. Other laws start with more basic global properties of the gas
in the galaxy, such as its total gas content.

Figures \ref{fig:1} - \ref{fig:4} show the radial distribution of
properties of the sample galaxies. The top panel shows observed
profiles of molecular, atomic, and total gas mass surface density. The
second and third panels show $v_{\rm circ}$ and $\beta$,
respectively. The observed and predicted $\Sigma_{\rm sfr}$ are shown
in the fourth panel.  Finally, the last panel shows ${\rm log}_{10}\:R_{\rm sfr}$. The
vertical dotted line in each plot shows the position of $r_{\rm out}$,
where atomic gas becomes dominant over molecular gas.

\begin{deluxetable}{ccccccccccccccc}
\tabletypesize{\footnotesize}
\tablecolumns{15}
\tablewidth{0pt}
\tablecaption{Star Formation Law Parameters}
\tablehead{\colhead{Galaxy} &
           \colhead{$r_{\rm out}$} &
           \colhead{$N_{\rm ann}$} &
           \colhead{$A_{g}$\tablenotemark{a}} &
           \colhead{$\chi_{g}$} &
           \colhead{$A_{\rm H2}$\tablenotemark{a}} &
           \colhead{$\chi_{\rm H2}$} &
           \colhead{$A_{\rm KM}$\tablenotemark{a}} &
           \colhead{$\chi_{\rm KM}$} &
           \colhead{$A_{\rm KMT}$\tablenotemark{a}} &
           \colhead{$\chi_{\rm KMT}$} &
           \colhead{$B_{\rm \Omega}$} &
           \colhead{$\chi_{\Omega}$} &
           \colhead{$B_{\rm CC}$} &
           \colhead{$\chi_{\rm CC}$} \\
	   \colhead{NGC:} &
	   \colhead{\tiny (kpc)} &
	   \colhead{} &
	   \colhead{\tiny($10^{-2}$)} &
	   \colhead{\tiny($10^{-2}$)} &
	   \colhead{\tiny($10^{-2}$)} &
	   \colhead{\tiny($10^{-2}$)} &
	   \colhead{} &
	   \colhead{\tiny($10^{-2}$)} &
	   \colhead{\tiny($10^{-2}$)} &
	   \colhead{\tiny($10^{-2}$)} &
	   \colhead{\tiny($10^{-3}$)} &
	   \colhead{\tiny($10^{-2}$)} &
	   \colhead{\tiny($10^{-3}$)} &
	   \colhead{\tiny($10^{-2}$)} 
}
\startdata
\multicolumn{15}{c}{Molecular Dominated regions} \\
\hline\noalign{\smallskip}
0628&4.5&20&7.75&9.55&4.90&12.6&0.737&32.6&3.13&11.6&3.63&24.4&4.65&15.9\\
2841&7.8&14&17.1&27.2&5.66&16.0&0.786&19.7&2.14&7.39&5.33&7.65&5.42&6.99\\
2903&5.2&19&6.67&17.3&5.56&21.6&1.92&22.5&4.24&20.2&6.85&20.1&9.99&23.9\\
3184&5.4&16&6.81&9.11&4.28&13.1&1.56&16.7&2.66&17.3&6.21&12.4&11.2&17.9\\
3351&5.6&18&12.2&21.8&5.75&24.3&1.02&19.0&2.85&32.5&5.52&10.0&6.67&19.3\\
3521&5.9&18&6.13&8.54&4.90&4.81&1.36&17.7&3.71&4.93&5.22&11.0&7.31&5.69\\
3627&8.1&28&10.0&15.7&6.35&17.4&2.29&24.1&3.97&26.1&9.12&22.4&11.3&16.2\\
4254&9.1&15&4.00&7.41&4.38&3.37&3.38&13.8&3.62&3.13&8.64&13.2&11.7&4.95\\
4321&8.8&20&4.46&17.5&3.75&14.2&1.82&24.0&2.63&10.1&5.90&21.9&7.90&16.3\\
4579&10.7&17&5.27&12.0&2.72&8.57&1.06&18.6&1.55&14.7&4.60&15.7&6.16&11.1\\
4736&2.0&14&8.54&12.3&7.83&18.9&0.690&28.0&5.96&26.8&3.22&24.8&3.58&20.8\\
5055&8.4&27&4.46&12.8&3.65&5.82&1.45&29.8&2.65&8.36&4.99&26.2&5.63&16.9\\
5194&6.9&28&4.10&21.2&4.15&14.9&1.46&29.7&3.22&12.8&4.70&29.0&5.53&22.7\\
5457&6.9&30&6.67&13.8&4.06&15.0&1.13&33.1&2.62&13.3&4.91&27.3&6.27&18.8\\
6946&6.0&33&5.96&27.0&5.74&21.9&2.49&32.3&4.31&17.9&7.81&30.7&10.8&23.6\\
7331&7.2&16&7.18&6.48&4.82&5.10&1.31&23.5&3.33&8.86&5.50&15.5&7.12&5.11\\
\hline\noalign{\smallskip}
$N_{\rm fit}$=16 &&333&&16.9&&15.4&&26.1&&16.8&&22.4&&17.5\\
$N_{\rm fit}$=1 &&333&6.55&22.9&4.74&18.2&1.45&31.2&3.17&20.5&5.66&24.8&7.28&22.4\\
\hline\hline\noalign{\smallskip}
\multicolumn{15}{c}{Molecular Rich regions} \\
\hline\noalign{\smallskip}
0628&5.6&5&7.04&8.67&6.35&5.32&2.10&5.15&3.52&5.75&6.18&7.13&6.22&7.20\\
2841&10.0&5&10.7&1.53&5.68&3.69&1.80&6.09&2.20&2.78&6.87&0.508&6.87&0.508\\
2903&5.7&2&8.43&2.21&6.79&4.44&2.82&4.01&3.59&2.92&8.45&1.32&8.95&1.05\\
3184&6.8&4&6.50&5.15&5.72&6.56&2.93&7.15&3.13&6.03&7.87&5.16&9.69&5.26\\
3351&7.5&6&7.93&6.30&3.90&10.1&1.50&11.5&1.53&8.59&5.90&7.01&5.91&6.99\\
3521&6.9&3&4.70&3.85&4.83&0.46&2.20&0.37&2.84&1.98&5.55&5.05&5.73&5.32\\
3627&8.4&1&15.0&-&5.24&-&2.10&-&1.56&-&9.43&-&9.47&-\\
4254&12.3&5&5.64&4.68&6.63&5.51&7.79&6.36&4.11&5.33&14.3&4.54&14.5&4.47\\
4321&11.0&5&5.50&5.93&4.28&4.99&2.86&4.62&2.02&4.42&7.35&4.59&7.46&4.59\\
4579&11.4&1&9.03&-&3.55&-&1.49&-&1.10&-&6.08&-&6.21&-\\
4736&4.2&15&7.92&4.17&4.13&7.73&0.811&7.61&1.68&12.4&3.64&6.01&3.64&6.01\\
5055&10.0&5&4.48&4.39&3.82&7.65&3.16&7.71&2.05&6.18&7.68&3.97&7.68&3.97\\
5194&7.9&4&4.68&6.47&4.03&3.91&2.15&3.43&2.20&4.54&5.83&6.04&5.84&6.05\\
5457&9.0&9&7.41&5.42&6.47&6.86&3.98&7.17&3.70&5.63&10.5&4.05&10.6&4.04\\
6946&7.5&8&13.2&12.1&12.2&12.0&7.27&13.0&6.80&10.1&18.3&11.8&18.8&11.5\\
7331&9.0&4&4.52&1.65&4.17&4.89&2.23&6.15&2.33&3.08&5.75&1.75&5.80&1.63\\
\hline\noalign{\smallskip}
$N_{\rm fit}$= 14&&80&&6.24&&7.40&&7.90&&8.00&&6.20&&6.16\\
$N_{\rm fit}$= 1&&82&7.12&14.9&5.37&16.3&2.41&31.5&2.67&20.7&7.25&21.9&7.38&22.3\\
\hline\hline\noalign{\smallskip}
\multicolumn{15}{c}{Combined Regions} \\
\hline\noalign{\smallskip}
0628&5.6&25&7.60&9.36&5.16&12.3&0.908&34.5&3.20&10.8&4.04&23.9&4.93&15.4\\
2841&10.0&19&15.1&24.9&5.67&13.7&0.977&23.6&2.15&6.44&5.70&8.20&5.77&7.56\\
2903&5.7&21&6.82&16.7&5.66&20.7&1.99&22.0&4.17&19.3&6.99&19.2&9.89&22.7\\
3184&6.8&20&6.75&8.39&4.54&13.0&1.77&18.8&2.75&15.8&6.51&12.0&10.9&16.3\\
3351&7.5&24&10.9&20.7&5.22&22.7&1.12&18.8&2.44&30.7&5.62&9.31&6.47&17.1\\
3521&6.9&21&5.90&8.98&4.89&4.44&1.46&17.9&3.57&6.18&5.27&10.3&7.06&6.69\\
3627&8.4&29&10.2&15.8&6.31&17.1&2.28&23.7&3.85&26.7&9.13&22.0&11.2&16.0\\
4254&12.3&20&4.36&9.46&4.86&8.90&4.17&20.2&3.74&4.38&9.80&15.0&12.4&6.27\\
4321&11.0&25&4.65&16.2&3.85&13.0&1.99&22.9&2.50&10.3&6.16&20.0&7.81&14.6\\
4579&11.4&18&5.43&12.9&2.76&8.75&1.08&18.4&1.52&14.7&4.68&15.5&6.16&10.7\\
4736&4.2&29&8.21&9.02&5.63&19.9&0.750&20.1&3.09&34.6&3.43&17.6&3.61&14.8\\
5055&10.0&32&4.46&11.8&3.67&6.04&1.64&30.2&2.54&8.97&5.34&25.0&5.91&16.3\\
5194&7.9&32&4.17&20.0&4.14&14.0&1.53&28.4&3.07&13.2&4.83&27.3&5.57&21.3\\
5457&9.0&39&6.83&12.5&4.53&16.0&1.51&37.3&2.84&13.5&5.85&27.8&7.08&19.2\\
6946&7.5&41&6.96&28.3&6.64&24.1&3.07&34.8&4.71&18.4&9.22&31.6&12.0&23.7\\
7331&9.0&20&6.55&10.1&4.69&5.56&1.46&23.0&3.10&10.2&5.55&13.8&6.83&5.88\\
\hline\noalign{\smallskip}
$N_{\rm fit}$=16 &&415&&16.5&&15.7&&26.8&&17.9&&21.5&&16.6\\
$N_{\rm fit}$=1 &&415&6.68&21.7&4.85&17.9&1.60&32.4&3.05&20.9&5.94&24.6&7.30&22.3\\
\enddata

\tablenotetext{a}{Units: $M_\odot {\rm yr^{-1}kpc^{-2}}$}

\label{tab:results}
\end{deluxetable}

Table~\ref{tab:results} shows the best-fit parameters of the star
formation laws as fit to the 16 sample galaxies, together with their
goodness of fit parameters, $\chi$. Results are shown separately for
the molecular dominated regions, the molecular rich regions with
$\Sigma_{\rm HI}/2<\Sigma_{\rm H2}<\Sigma_{\rm HI}$ and the combined
regions.
For molecular dominated regions, when each galaxy is allowed one free
parameter, the Constant Molecular law is the most accurate model,
followed by the KMT09 law, the Kennicutt-Schmidt law, the GMC
Collision law, the Gas-$\Omega$ law, and the KM05 law. But the first
four have similar values of $\chi$: the dispersion in the residuals of
the GMC Collision law is a factor of 1.50, while that of the Constant
Molecular law is 1.43. Even the worst-fitting relation, i.e. the KM05
law has a dispersion of only 1.82.  The Constant Molecular law is
still the best-fitting model when only one free parameter is allowed
for the whole sample (with rms error of factor of 1.52), followed by
the KMT09 law, the GMC collision law (rms error factor of 1.67), the
Kennicutt-Schmidt law, the Gas-$\Omega$ law, and the KM05 law (rms
error factor of 2.05).



For the extended molecular rich regions with $\Sigma_{\rm
  HI}/2<\Sigma_{\rm H2}<\Sigma_{\rm HI}$, all the star formation laws
still work reasonably well. When each galaxy with $N_{\rm ann}\geq3$
in such regions (note, NGC 3627 and NGC 4579 have $N_{\rm ann}=1$, so
the analysis is not performed on them individually) has one free
parameter, the ordering of the laws from best to worst is GMC
Collision, Gas-$\Omega$, Kennicutt-Schmidt, Constant Molecular, KM05,
KMT09. However, again, the differences are relatively minor and a
different ordering results when allowing only a single global free
parameter (see Table~\ref{tab:results}).


Finally, the same analysis is repeated for the combined molecular
dominated and rich regions (Table~\ref{tab:results}). The Constant
Molecular law gives the best fit with rms error of a factor of 1.4 for
the case of one free parameter per galaxy and 1.5 for the case of one
global free parameter followed by the Kennicutt-Schmidt law, GMC
Collision law, KMT09 law, Gas-$\Omega$ law, and KM05 law, which have
rms error of factor of 1.8.

We summarize the rms dispersion of the fitted star formation law
residuals, ${\rm log}_{10}\: R_{\rm sfr}$, in Table~\ref{tab:rms}. We
attribute the smaller dispersion in the molecular rich regions to
these being relatively narrow annular parts of galaxies that do not
span a wide range of galactic properties, such as gas mass surface
densities and orbital velocities. As described above, the differences
between the different star formation laws are relatively modest. For
molecular dominated or entire regions, there is a range from about a
factor of 1.4 to 1.8 dispersion when allowing each galaxy one free
parameter to normalize the star formation law, rising to 1.5 to 2.1
when a single global parameter is fit to the sample. 

In Table~\ref{tab:rms} we also show the effect of excluding the
central $r<$1~kpc regions of the galaxies on the rms dispersions:
these results are shown in parentheses. These central regions have
poorly resolved rotation curves and have more uncertain molecular gas
masses (Sandstrom et al. 2013; see also Bell et al. 2007; Israel et
al. 2009). Excluding these regions always leads to a reduction in
log$_{10}\:R_{\rm sfr}$, typically by $\sim 10-20\%$, and can result
in occasional changes of the relative ordering of the different star
formation laws. However, generally the Constant Molecular,
Kennicutt-Schmidt, KMT09 and GMC Collision laws are seen to give quite
similar rms values that are smaller than those of the KM05 and
Gas-$\Omega$ laws. So we conclude that the results of the main
analysis are not being adversely affected by the central kpc regions.

As described at the start of this section, the relative values of the
dispersion between the different laws can result not only from the
intrinsic merit of the physical model, but also because of the varying
types of inputs and their associated observational uncertainties. The
KMT09 law is an extension of the KM05 model, and it is seen to give
improved fits to the data, i.e., with smaller dispersions. Likewise,
the GMC Collision model can be regarded as a modification and
extension of the dynamical Gas-$\Omega$ model: it uses the same inputs
plus an additional one, the gradient of the rotation curve. This
generally leads to a modest reduction of the dispersions of the data
sets that include the molecular dominated regions where the largest
rotation curve gradients are present (see also
\S\ref{S:beta}). The more empirical Kennicutt-Schmidt and
Constant Molecular laws also provide good fits to the data, with the
latter giving the (modestly) best fit for the molecular dominated and
entire regions. However, this may reflect its more limited physical
scope of starting with the observed amount of molecular gas as its
input, rather than trying to connect star formation activity to more
fundamental galactic properties.


\begin{deluxetable}{ccccccc}
\tabletypesize{\footnotesize}
\tablecolumns{7}
\tablewidth{0pt}
\tablecaption{RMS dispersion of star formation law residuals (${\rm log}_{10}\: R_{\rm sfr}$)\tablenotemark{a} 
}
\tablehead{
\colhead{} &
\colhead{N$_{\rm fit}$ = 16} &
\colhead{N$_{\rm fit}$ = 14} &
\colhead{N$_{\rm fit}$ = 16} &
\colhead{} &
\colhead{N$_{\rm fit}$ = 1} &
\colhead{}\\
\hline
\noalign{\smallskip}
\colhead{Star Formation Law} &
\colhead{Molecular} &
\colhead{Molecular} &
\colhead{Entire} &
\colhead{Molecular} &
\colhead{Molecular} &
\colhead{Entire}\\
\colhead{} &
\colhead{Dominated} &
\colhead{Rich} &
\colhead{Regions} &
\colhead{Dominated} &
\colhead{Rich} &
\colhead{Regions}
}
\startdata

Kennicutt-Schmidt &     0.165 (0.149)&     0.0567&     0.161 (0.144)&     0.229 (0.217)&     0.151&     0.217 (0.204)\\
Constant Molecular &    0.150 (0.131)&     0.0672&     0.153 (0.135)&     0.181 (0.164)&     0.161&     0.179 (0.166)\\
KM05 &                        0.254 (0.228)&     0.0718&     0.263 (0.237)&     0.312 (0.285)&     0.310&     0.323 (0.301)\\
KMT09 &                      0.164 (0.134)&     0.0727&     0.176 (0.142)&     0.205 (0.184)&     0.209&     0.208 (0.192)\\
Gas-$\Omega$ &           0.218 (0.191)&     0.0563&     0.211 (0.182)&     0.248 (0.219)&     0.215&     0.246 (0.220)\\
GMC Collision &            0.171 (0.146)&     0.0559&     0.163 (0.140)&     0.223 (0.193)&     0.219&     0.222 (0.199)\\


\enddata
\tablenotetext{a}{Results in parentheses show the effect of excluding the $r<1$~kpc regions}
\label{tab:rms}
\end{deluxetable}

\subsection{Dependence of Star Formation Law Parameters on Galactic Dynamical Properties}\label{S:properties}

\begin{deluxetable}{ccccccccccccccc}
\tabletypesize{\footnotesize}
\tablecolumns{15}
\tablewidth{0pt}
\tablecaption{Star formation law parameters for galactic dynamical property sub-samples}
\tablehead{\colhead{} &
           \colhead{} &
           \colhead{$N_{\rm ann}$} &
           \colhead{$A_{g}$\tablenotemark{a}} &
           \colhead{$\chi_{g}$} &
           \colhead{$A_{\rm H2}$\tablenotemark{a}} &
           \colhead{$\chi_{\rm H2}$} &
           \colhead{$A_{\rm KM}$\tablenotemark{a}} &
           \colhead{$\chi_{\rm KM}$} &
           \colhead{$A_{\rm KMT}$\tablenotemark{a}} &
           \colhead{$\chi_{\rm KMT}$} &
           \colhead{$B_{\rm \Omega}$} &
           \colhead{$\chi_{\Omega}$} &
           \colhead{$B_{\rm CC}$} &
           \colhead{$\chi_{\rm CC}$} \\
	   \colhead{} &
	   \colhead{} &
	   \colhead{} &
	   \colhead{\tiny($10^{-2}$)} &
	   \colhead{\tiny($10^{-2}$)} &
	   \colhead{\tiny($10^{-2}$)} &
	   \colhead{\tiny($10^{-2}$)} &
	   \colhead{} &
	   \colhead{\tiny($10^{-2}$)} &
	   \colhead{\tiny($10^{-2}$)} &
	   \colhead{\tiny($10^{-2}$)} &
	   \colhead{\tiny($10^{-3}$)} &
	   \colhead{\tiny($10^{-2}$)} &
	   \colhead{\tiny($10^{-3}$)} &
	   \colhead{\tiny($10^{-2}$)} 
}
\startdata
Low $\bar{\beta}$&$N_{\rm fit}$ = 9&195&&16.5&&15.2&&27.9&&18.2&&23.3&&17.4\\
&$N_{\rm fit}$= 1&195&7.34&24.4&4.91&17.8&1.21&31.6&0.03&20.7&5.16&25.9&6.19&21.4\\

High $\bar{\beta}$&$N_{\rm fit}$= 7&138&&17.5&&15.6&&23.3&&14.6&&20.9&&17.6\\
&$N_{\rm fit}$= 1&138&5.58&18.7&4.50&18.5&1.87&27.2&3.21&20.4&6.45&22.2&9.17&19.7\\

\hline\noalign{\smallskip}
KS Test $p$ &&&0.032&&0.678&&0.067&&0.620&&0.067&&0.011&\\

\hline\noalign{\smallskip}

Low $\bar{v}$&$N_{\rm fit}$= 7&143&&18.7&&19.0&&24.5&&22.3&&21.9&&19.6\\
&$N_{\rm fit}$= 1&143&7.39&23.1&5.65&20.0&1.83&31.2&3.86&23.9&6.84&25.1&9.22&24.6\\

High $\bar{v}$&$N_{\rm fit}$= 9&190&&15.5&&11.9&&27.2&&10.9&&22.7&&15.8\\
&$N_{\rm fit}$= 1&190&5.98&22.1&4.14&14.1&1.22&29.1&2.73&14.6&4.91&22.8&6.10&16.8\\

\hline\noalign{\smallskip}
KS Test $p$&&&0.735&&0.067&&0.067&&0.067&&0.011&&0.017&\\

\hline\noalign{\smallskip}

Non-barred&$N_{\rm fit}$= 7&134&&15.6&&12.0&&27.1&&12.6&&22.9&&16.2\\
&$N_{\rm fit}$= 1&134&6.13&25.1&4.69&15.1&1.24&33.5&3.20&16.7&4.86&25.2&5.78&20.7\\

Barred&$N_{\rm fit}$= 9&199&&17.7&&17.2&&25.4&&19.1&&22.0&&18.4\\
&$N_{\rm fit}$= 1&199&6.85&21.2&4.77&20.0&1.61&28.7&3.15&22.8&6.27&23.6&8.50&20.9\\

\hline\noalign{\smallskip}
KS Test $p$&&&0.620&&0.996&&0.358&&0.790&&0.155&&0.017&\\
\enddata

\tablenotetext{a}{Units: $M_\odot {\rm yr^{-1}kpc^{-2}}$}

\label{tb:4}
\end{deluxetable}

We test the dependence of the derived star formation law parameters
with three basic galactic dynamical properties: (1) Galactic Disk Shear, as measured by the gradient of the rotation
curve; (2) Rotation Speed; (3) Presence of a Bar.

\subsubsection{Galactic Disk Shear}\label{S:beta}

\begin{figure}[!tb]
\centering
\begin{tabular}{cc}
\epsfig{file=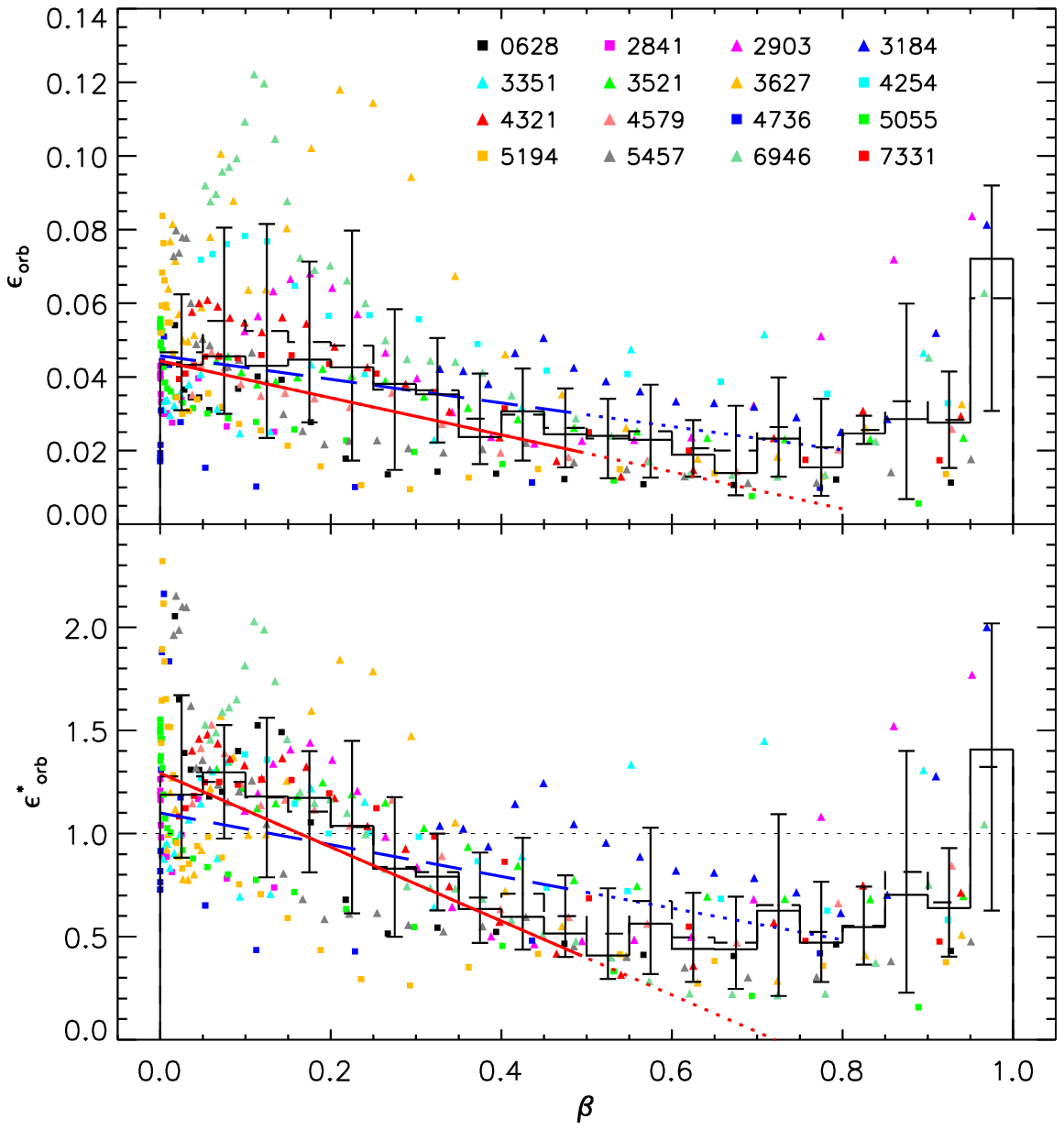,width=0.45\linewidth,clip=} 
\epsfig{file=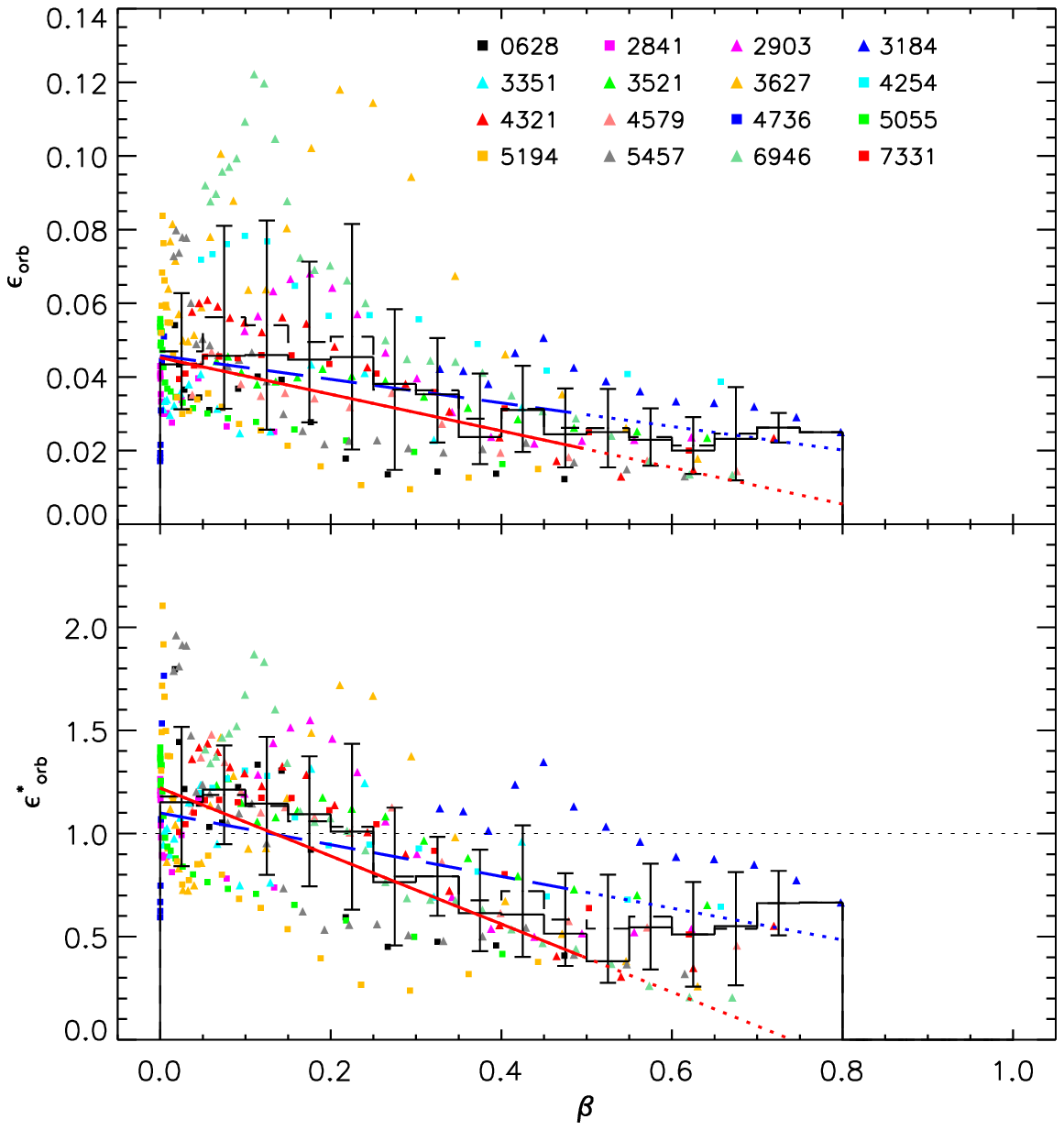,width=0.45\linewidth,clip=} 
\end{tabular}
\caption{\label{fig:beta} 
(a) Top left: Star formation efficiency per orbital time,
  $\epsilon_{\rm orb}$, as a function of rotation curve gradient,
  $\beta$.
The data for the annuli in each galaxy are shown with different colors
and symbols, as indicated.  Also shown are the mean (black dashed) and
median (black solid) of the data, together with $1\sigma$ dispersion,
in uniform bins of $\beta$. The best fit linear relation
$\epsilon_{\rm orb} = \epsilon_{\rm orb,0} (1.0-\alpha_{\rm cc}\beta)$
for data in the range $0<\beta<0.5$ (see text for assumed errors) is
shown by the solid red line (with extrapolation for $\beta>0.5$ shown
by a dotted line). The best fit linear relation with $\alpha_{\rm
  cc}=0.7$ from the GMC collision theory (for $\beta\ll1)$ is shown by
the dashed blue line (with extrapolation for $\beta>0.5$ shown
again by a dotted line).
(b) Bottom left: As above, but now showing normalized efficiency per orbit, $\epsilon^*_{\rm
  orb} \equiv \epsilon_{\rm orb}/\bar{\epsilon}_{\rm orb}$ versus
$\beta$. Each value has been normalized by the average for its
particular galaxy.
The best fit line for $\epsilon^*_{\rm orb} = \epsilon^*_{\rm
  orb,0} (1.0-\alpha_{\rm cc}\beta)$ for data in the range
$0<\beta<0.5$ is shown by the solid red line, and that for
$\alpha_{\rm cc}=0.7$ by the dashed blue line (with extrapolations for $\beta>0.5$ shown
by dotted lines).  
(c) Top right: Same as (a), but now excluding all data at $r<1$~kpc.
(d) Bottom right: Same as (b), but now excluding all data at $r<1$~kpc.
In all panels, a trend of declining efficiency with increasing
$\beta$ (i.e. decreasing shear rate) is seen: there is about a factor
of two decrease as $\beta$ rises from 0 (flat rotation curve case) to
0.5. 
}
\end{figure}

We first divide the galaxy sample into ``Low $\bar{\beta}$'' and
``High $\bar{\beta}$'' sub-samples using a boundary of
$\bar{\beta}=0.3$. There are 9 Low $\bar{\beta}$ galaxies and 7 High
$\bar{\beta}$ galaxies. We repeat the analysis of \S\ref{S:lawtest}
for these two sub-samples and the results are shown in
Table~\ref{tb:4}.

We carry out a two-sample Kolmogorov-Smirnov (KS) test to see if the
distribution of derived star formation law parameters of the
individual galaxies in each sub-sample, e.g., $A_g$, $A_{\rm H2}$,
etc., are consistent with being drawn from the same parent
distribution. The probabilities that they do come from the same
distribution are also shown in Table~\ref{tb:4}. In general, none of
the star formation laws show very significant probabilities for
systematic differences between the Low and High $\bar{\beta}$
sub-samples, in part because of the small numbers in these
samples. 

The most significant of these is the approximately 1\% probability
that the $B_{\rm CC}$'s from the GMC Collision model are drawn from
the same distribution. This could be explained by the fact that because the
equation for the star formation rate (eq.~\ref{sfrcoll}) depends on
both $B_{\rm CC}$ and $\beta$, these quantities become (inversely)
correlated: i.e., a galaxy with high $\bar{\beta}$ tends to need a
larger value of $B_{\rm CC}$ to yield a given star formation
rate. Given the relatively low significance of the probability, little
more can be concluded for these trends with mean galactic shear until
larger samples of galaxies are available.

We can, however, look in more detail at each galactic annulus and test
for the prediction of the GMC Collision law that the star formation
efficiency per orbit, $\epsilon_{\rm orb}=(2\pi/\Omega)\Sigma_{\rm
  sfr}/\Sigma_g$, should decline as the {\it local} value of $\beta$
in a given annulus increases. In the Gas-$\Omega$ model,
$\epsilon_{\rm orb} = 2\pi B_\Omega = {\rm constant}$. In the GMC
Collision model, $\epsilon_{\rm orb} = 2\pi B_{\rm
  CC}Q^{-1}(1-0.7\beta)$ (for $\beta\ll 1$). The star formation rate
and efficiency decline with increasing $\beta$ because a lower shear
rate leads to a smaller rate of GMC collisions.

To test for this effect, in Figure~\ref{fig:beta} we show
$\epsilon_{\rm orb}$ versus $\beta$ for each annulus (data from each
galaxy are shown with different colors and symbols) together with the
mean, median and $1\sigma$ dispersion in binned intervals of
$\beta$. We also show a graph of $\epsilon^*_{\rm orb} \equiv
\epsilon_{\rm orb}/\bar{\epsilon}_{\rm orb}$, i.e., where each value
has been normalized by the average for its particular galaxy. Versions
of these graphs where all data at $r<1$~kpc have been excluded are
also shown.

A trend of declining efficiency with increasing $\beta$ is seen: there
is about a factor of two decrease as $\beta$ rises from 0 (flat
rotation curve case) to 0.5. There is a flattening and hint of an
upturn in $\epsilon_{\rm orb}$ and $\epsilon^*_{\rm orb}$ as $\beta$
reaches 1 (solid body rotation), but there are very few data points
near $\beta=1$ and, as can be seen from Figure~\ref{fig:beta}c and d,
those that exist are all located at galactic centers, where there may
be larger systematic uncertainties, e.g., in determining the rotation
curve shape (perhaps producing overestimated values of $\beta$ in
regions where the rotation curve is insufficiently resolved).  On the
other hand, the main trend of declining efficiency with increasing
$\beta$ is not driven by the presence of the very central regions: we
further test for this by excluding data from the central 1, 2, 3~kpc
and find essentially the same results for $\beta<0.8$. This also
indicates that the decline in $\epsilon^*_{\rm orb}$ is not being
driven by a systematic change in the ``X-factor'' that is needed to
estimate $\Sigma_{\rm H2}$ from observed CO line intensity, since
Sandstrom et al. (2013) find this conversion factor is constant in
these galaxies in all but the very inner $\sim$1 kpc regions.

To gauge the decline of star formation efficiency per orbit with
$\beta$ more quantitatively and to assess the significance of this
trend, for the data in the range $0<\beta<0.5$, we derive the best-fit
function $\epsilon_{\rm orb} = \epsilon_{\rm orb,0} (1 - \alpha_{\rm
  CC} \beta)$, finding $\epsilon_{\rm orb,0} 
= 0.044\pm0.005$ and $\alpha_{\rm CC}=1.13\pm 0.49$ when using data
that include the galactic centers and $\epsilon_{\rm orb,0} =
0.045\pm0.005$ and $\alpha_{\rm CC}=1.10\pm 0.44$ when excluding data
at $r<1$~kpc (with the errors based on an assumption of 50\% typical
uncertainties in the absolute values of $\Sigma_{\rm sfr}$ and
$\Sigma_g$ and a 20\% uncertainty in $\Omega$, yielding $73\%$
uncertainty in $\epsilon_{\rm orb}$ to which we also add a minimum
threshold uncertainty equal to the observed standard deviation of
0.022, and an assumed $30\%$ plus threshold of 0.14 uncertainty in
$\beta$). This best fit function is shown by the red solid line in
Figure~\ref{fig:beta}. Note, this line tends to sit below the binned
mean and median values because the assumed errors in $\epsilon_{\rm
  orb}$ have a component that is proportional to $\epsilon_{\rm orb}$.

Similarly for $\epsilon^*_{\rm orb}$ we derive $\epsilon^*_{\rm
  orb,0}=1.29\pm 0.08$ and $\alpha^*_{\rm CC}=1.39\pm 0.32$ when using
data that include the galactic centers and $\epsilon^*_{\rm
  orb,0}=1.22\pm 0.08$ and $\alpha^*_{\rm CC}=1.35\pm 0.31$ when
excluding data at $r<1$~kpc (with the errors based on an assumption of
25\% typical uncertainties in the relative (disk-normalized) values of
$\Sigma_{\rm sfr}$ and $\Sigma_g$ and a 10\% uncertainty in relative
values of $\Omega$, yielding $37\%$ uncertainty in $\epsilon^*_{\rm
  orb}$ to which we also add a minimum threshold uncertainty equal to
the observed standard deviation of 0.40, and an assumed $30\%$ plus
threshold of 0.14 uncertainty in $\beta$). By this measure, a
dependence of $\epsilon^*_{\rm orb}$ on $\beta$ (i.e., a non-zero
value of $\alpha_{\rm cc}$) is detected at about the $4\sigma$ level,
although the precise level of this significance is dependent on the
rather uncertain assumptions about the size of the
uncertainties. 

Evaluating the Spearman rank correlation coefficient, $r_s$, and
probability for chance correlation, $p_s$, for these data (i.e., for
$0<\beta<0.5$), we find $r_s = -0.27$ and $p_s=1.1\times 10^{-5}$ for
$\epsilon_{\rm orb}$ versus $\beta$ (essentially the same values are
found if the $r<1$~kpc data are excluded) and $r_s = -0.49$ and
$p_s=4.8\times 10^{-17}$ for $\epsilon^*_{\rm orb}$ versus $\beta$
($r_s = -0.44$ and $p_s=1.6\times 10^{-13}$ are found if the $r<1$~kpc
data are excluded), which suggests we may have been too conservative
in our estimates of the uncertainties. Thus we conclude there is
strong evidence of declining star formation efficiency per orbit with
increasing rotation curve gradient $\beta$ (i.e. declining shear).


Such a decline in star formation efficiency with increasing $\beta$,
i.e., decreasing shear rate, is the opposite of what would be expected
if formation of star-forming clouds (i.e., GMCs) from the diffuse
interstellar medium via gravitational instability was the rate
limiting step for galactic star formation rates, since increasing
shear acts to stabilize gas disks. However, such a decline is
predicted by the GMC Collision model (formally with $\alpha_{\rm
  CC}\simeq 0.7$) for galactic star formation rates, where the rate limiting
step for star formation is formation of star-forming clumps within
GMCs via shear-driven GMC-GMC collisions (Tan
2000). Figure~\ref{fig:beta} also shows the predicted $\epsilon_{\rm
  orb}$ and $\epsilon^*_{\rm
  orb}$ versus $\beta$ relation (i.e., based on eq.~\ref{sfrcoll})
from the GMC Collision model for the range $0<\beta<0.5$ (note, this
model was developed for $\beta \ll 1$): it provides a reasonable match
to the data, although with a somewhat shallower slope $\alpha_{\rm CC}$.

\subsubsection{Rotation Speed}

\begin{figure}[!tb]
\centering
\begin{tabular}{cc}
\epsfig{file=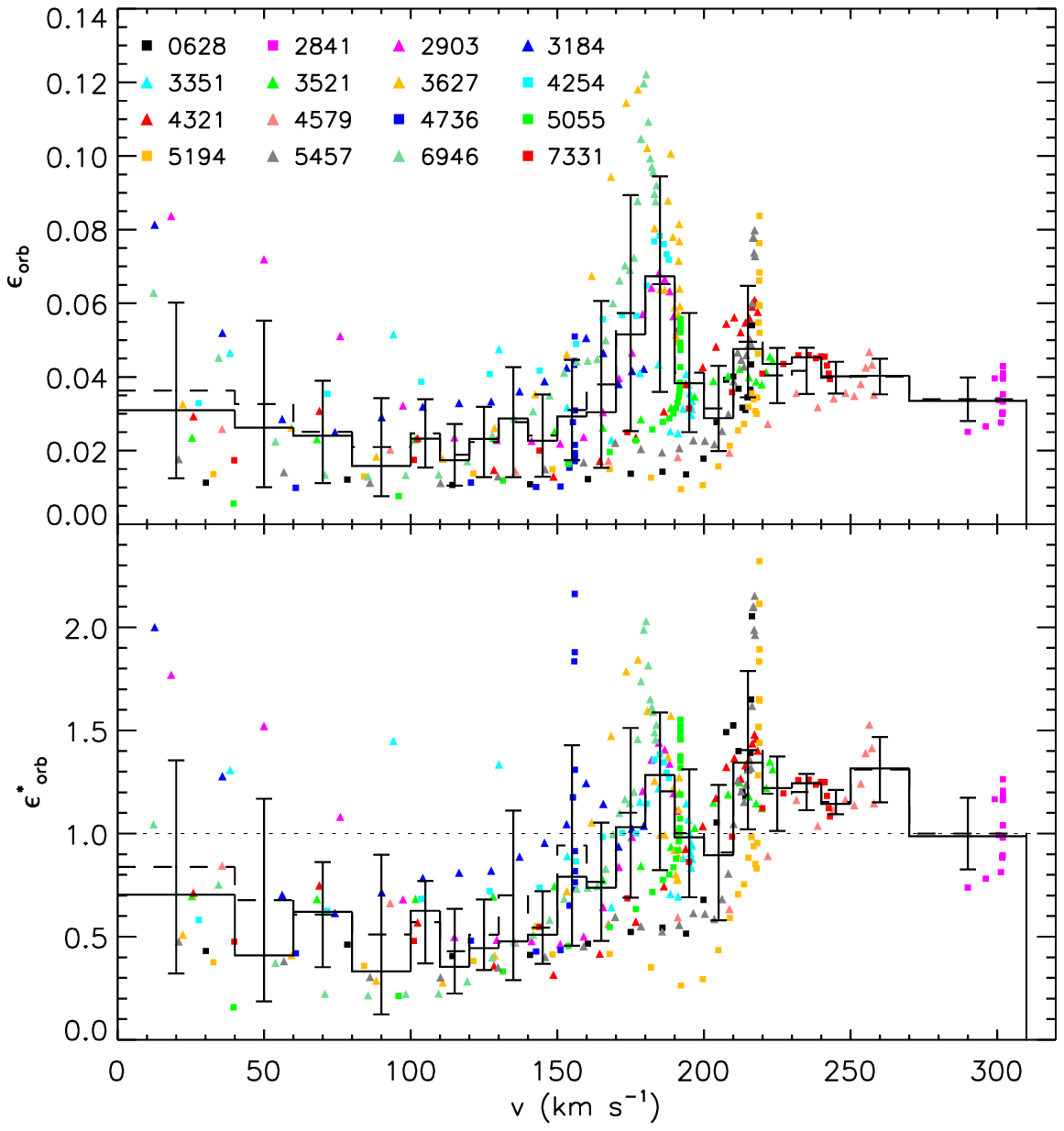,width=0.45\linewidth,clip=}
\epsfig{file=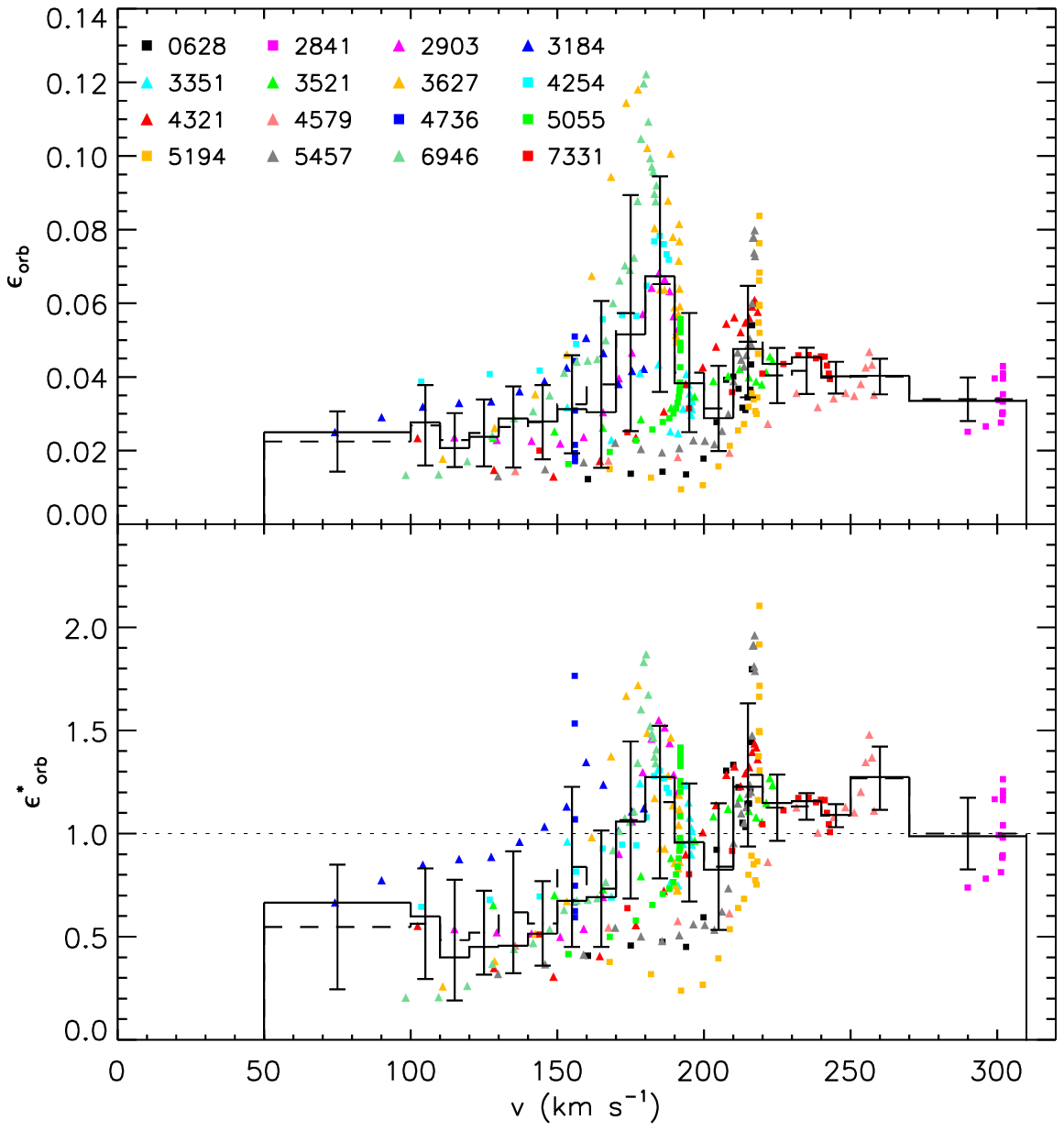,width=0.45\linewidth,clip=}
\end{tabular}

\caption{\label{fig:v} 
(a) Top left: Star formation efficiency per orbital time,
  $\epsilon_{\rm orb}$, as a function of rotation velocity, $v$.  The
  data for the annuli in each galaxy are shown with different colors
  and symbols, as indicated.  Also shown are the mean (dashed line)
  and median (solid line) of the data, together with $1\sigma$
  dispersion, in uniform bins of $v$.
(b) Bottom left: Normalized efficiency per orbit, $\epsilon^*_{\rm
  orb} \equiv \epsilon_{\rm orb}/\bar{\epsilon}_{\rm orb}$ versus
$v$. Each value has been normalized by the average for its particular
galaxy. 
(c) Top right: Same as (a), but now excluding all data at $r<1$~kpc.
(d) Bottom right: Same as (b), but now excluding all data at $r<1$~kpc.
In all panels, a trend of increasing efficiency with increasing $v$ up to
$v\simeq 170\:{\rm km\:s^{-1}}$ is seen (see text).
}
\end{figure}

We divide the galaxy sample into ``Low $\bar{v}$'' and ``High
$\bar{v}$'' sub-samples with $\bar{v}=173\:{\rm km\:s^{-1}}$ being the
dividing line. There are 7 Low $\bar{v}$ galaxies and 9 High
$\bar{v}$ galaxies. We repeat the analysis of \S\ref{S:lawtest}
for these two sub-samples and the results are shown in
Table~\ref{tb:4}.

We again carry out a two-sample Kolmogorov-Smirnov (KS) test to see if
the distribution of derived star formation law parameters of the
individual galaxies in each sub-sample are consistent with being drawn
from the same parent distribution. The probabilities that they do come
from the same distribution are shown in Table~\ref{tb:4}. As with the
similar analysis for galactic disk shear, there are no especially
significant differences between the sub-samples. Both ``dynamical''
star formation laws, i.e., Gas-$\Omega$ and GMC Collision that
involve galactic rotation as an input, show potential differences in
the sub-samples at the $1 - \sim 0.01$ probability. 
For this sample, Low $\bar{v}$ galaxies show higher star formation
efficiency per mean orbital time, similar to results reported by Leroy
et al. (2013). However, such a trend is expected from simple
correlated uncertainties, since, other things being equal, high
velocity systems will tend to have shorter orbital times. So to
explain a given star formation rate, a higher efficiency per orbit is
needed.

We investigate the dependence of star formation efficiency per orbit
with local $v$ in a given annulus in Figure~\ref{fig:v}. A trend of
increasing efficiency with increasing $v$ is seen. This is consistent
with the results of \S\ref{S:beta} showing declining efficiency with
increasing $\beta$, since high $\beta$ regions tend to have low $v$
(being near galactic centers). 

Note, this trend of increasing efficiency with increasing local $v$ is
the opposite of the trend with $\bar{v}$, discussed above. Such
opposite behavior was also seen for the dependence of $\epsilon_{\rm
  orb}$ on $\beta$ and $\bar{\beta}$. This may indicate that the
trends in galaxy averages, which are based on just 16 data points and
which do not span a very wide dynamic range, are being driven by
correlated uncertainties in $\epsilon_{\rm orb}$ and ($\bar{\beta}$,
$\bar{v}$). We expect that the more reliable indicator of the effect
on star formation efficiency per orbit of these galactic properties is
that shown by $\epsilon_{\rm orb}$ and $\epsilon^*_{\rm orb}$ versus
local values of $\beta$ and $v$, since they are based on a larger
number of independent data points that span a wider dynamic range.

Figure \ref{fig:v} shows a relatively constant average value of
$\epsilon_{\rm orb}\simeq 0.04$ at velocities $\gtrsim 170\:{\rm
  km\:s^{-1}}$. This indicates that these mostly flat rotation curve
galactic star-forming disks can be treated as self-similar systems,
turning a small, fixed fraction of their local total ($\rm H_2$ + \ion{H}{1}) gas content to stars every
local orbit, as described in both the Gas-$\Omega$ and GMC Collision
models.

\subsubsection{Presence of a Bar}

To test the effects on derived star formation law parameters of the
presence of a bar we make the following division of the main galaxy
sample. The non-barred sub-sample (with 7 galaxies) contains only
normal spiral galaxies (SAa - Sac). The barred sub-sample (with 9
galaxies) contains both barred type galaxies (SBb) and transition type
galaxies (SABb - SABbc).

As above, we carry out a two-sample Kolmogorov-Smirnov (KS) test to
see if the distribution of derived star formation law parameters of
the individual galaxies in each sub-sample are consistent with being
drawn from the same parent distribution. The probabilities that they
do come from the same distribution are shown in Table~\ref{tb:4}.
Again, there are no especially significant differences between the
sub-samples. The GMC Collision model shows a potential difference in
the sub-samples at the $1 - \sim 0.01$ probability, with barred
galaxies having a larger average value of $B_{\rm CC}$ (i.e. higher
star formation efficiency per orbit) than non-barred galaxies by about
50\%. In the context of this model, this might indicate an enhancement
in GMC-GMC collision rates with the presence of a bar (which is likely
also correlated with the presence of spiral arms in the main
star-forming disk; orbit crowding in spiral arms may lead to enhanced
GMC collision rates, e.g., Dobbs 2013), but a larger sample of
galaxies is needed to be able to test the significance of this
potential effect. We note on the other hand that Meidt et al. (2013)
have claimed there is actually a suppression of star formation (longer
molecular gas depletion times) in the spiral arms of M51 due to
enhanced streaming motions.

On fitting the Gas-$\Omega$ law, a modestly higher (by a factor of
about 1.3) star formation efficiency per orbit, $\epsilon_{\rm
  orb}=2\pi B_\Omega$, is also seen in the barred compared to
non-barred galaxies. However, the KS probability of the samples having
the same intrinsic distributions of efficiency parameters is a
relatively large 0.16, i.e., the effect is not very significant. This
result is consistent with that noted in the study of a larger sample
of more distant, less well-resolved galaxies by Saintonge et
al. (2012), who found a factor of 1.5 enhancement in molecular gas
depletion rates ($\propto \Sigma_{\rm sfr}/\Sigma_{\rm H2}$) in their
barred sample compared to their control sample (but with an even
larger KS probability of 0.25 of the samples being the same).

Finally we note that there are correlations amongst the properties of
the sub-samples: e.g., most barred galaxies are also high $\bar{\beta}$
galaxies. Thus one needs to be careful in attributing primary cause of
an effect on star formation to these dynamical properties.

\section{Discussion and Conclusions}\label{S:conclusion}

We have tested six star formation laws against the resolved profiles
of 16 molecular dominated and molecular rich regions of nearby,
massive disk galaxies. There is a range from about a factor of 1.4 to
1.8 dispersion in the residuals of the best-fits when allowing each
galaxy one free parameter to normalize the star formation laws, rising
to 1.5 to 2.1 when a single global parameter is fit to the sample for
each law.

Since the different laws involve different inputs, which can have
varying levels of observational uncertainties and varying degrees to
which they connect to fundmental galactic physical properties, the
relative ordering of the laws is not of primary importance (formally,
the Constant Molecular law does best in having the smallest residuals;
see Table 3). 

More interesting is the comparison of laws within similar
classes. Thus the turbulence-regulated model of KMT09 is seen to be a
clear improvement over the KM05 model. The GMC Collision model
improves over the Gas-$\Omega$ model. 

The reason for this latter effect is the predicted decrease in star
formation efficiency per orbital time with decreasing shear rate
(increasing $\beta$) in the disk due to a reduced rate of shear-driven
GMC-GMC collisions (Gammie et al. 1991; Tan 2000; Tasker \& Tan 2009),
which is elucidated in Figure~\ref{fig:beta}. We estimate that the
significance of this trend over the range $0<\beta<0.5$ is at least
at the $4\sigma$ level. Such a trend is the opposite of that expected if
development of gravitational instabilities (e.g., leading to GMC
formation) from the diffuse interstellar medium is the rate limiting
step for star formation activity.

Confirmation of this result with a larger sample of galaxies, together
with more careful investigation of potential systematic uncertainties,
such as galactic radial gradients in normalization of star formation
rate indicators and CO to $\rm H_2$ conversion factors (although this
latter does not appear to be a major effect; Sandstrom et al. 2013),
is desirous.

More tentatively, we have found evidence that the presence of a bar
boosts star formation efficiency per orbit. This could potentially be
due to the influence of the bar on the strength of spiral arms (or
more general axisymmetric structure; Kendall et al. 2011) in the
larger-scale star-forming disks of the galaxies, although the
influence of spiral arms on star formation activity in NGC 628, NGC
5194 and NGC 6946 has been found to be small ($\lesssim 10\%$) (Foyle
et al. 2010). More detailed study of the influence of spiral arms
(including potential inducement by the presence of bars) and their
effect on star formation efficiency per orbit in a larger sample of
galaxies is needed. Also worthwhile is further theoretical work on the
influence of bars and spiral arms on the global GMC collision rate and
its link to star formation, i.e., compared to that in more
axisymmetric, flocculent galaxies.

\acknowledgements CS acknowledges support from the Royal Thai Government Scholarship.
JCT acknowledges NASA Astrophysics Theory and
Fundamental Physics grant ATP09-0094. This research has made use of
the NASA/IPAC Extragalactic Database (NED), which is operated by the
Jet Propulsion Laboratory, California Institute of Technology, under
contract with the National Aeronautics and Space Administration.


\end{document}